\documentclass[twocolumn]{article}

\usepackage{arxiv}

\usepackage[utf8]{inputenc} 
\usepackage[T1]{fontenc}    
\usepackage{hyperref}       
\usepackage{url}            
\usepackage{booktabs}       
\usepackage{amsfonts}       
\usepackage{nicefrac}       
\usepackage{microtype}      
\usepackage{cleveref}       
\usepackage{lipsum}         
\usepackage{graphicx}
\usepackage{doi}
\usepackage{tikz}
\usetikzlibrary{positioning,arrows.meta,calc,shapes, shapes.geometric}

\usepackage{multirow}
\usepackage{longtable}
\usepackage{float}

\usepackage[
  style=numeric-comp,
  datamodel=software, 
  abbreviate=false,
  natbib=true,
  sorting=none,
  backend=bibtex,
  bibencoding=utf8,
  url=true,
  doi=true,
  defernumbers,
  maxcitenames=3,
  defernumbers=false,
  maxbibnames=100]{biblatex}
\bibliography{references}

\definecolor{myColor}{rgb}{0,0.1,0.55}

\title{5G NR PRACH Detection with Convolutional Neural Networks (CNN): Overcoming Cell Interference Challenges}

\newif\ifuniqueAffiliation
\uniqueAffiliationtrue

\ifuniqueAffiliation 
\author{ \href{https://orcid.org/0000-0001-9612-0266}{\includegraphics[scale=0.06]{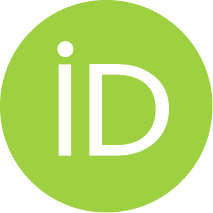}\hspace{1mm}Desire~Guel}\thanks{Use footnote for providing further
		information about author (webpage, alternative
		address)---\emph{not} for acknowledging funding agencies.} \\
	Department of Computer Science\\
	Joseph KI-ZERBO University\\
	Ouagadougou, Burkina Faso \\
	\texttt{desire.guel@ujkz.bf} \\
	\And
	Arsène~Kaboré \\
	Department of Computer Science\\
	Joseph KI-ZERBO University\\
	Ouagadougou, Burkina Faso \\
	\texttt{arsenekabore670@gmail.com} \\
	\And
	\href{https://orcid.org/0000-0002-0629-5888}{\includegraphics[scale=0.06]{orcid.pdf}\hspace{1mm}Didier~Bassolé} \\
	Department of Computer Science\\
	Joseph KI-ZERBO University\\
	Ouagadougou, Burkina Faso \\
	\texttt{didier.bassole@ujkz.bf} \\
}
\else
\usepackage{authblk}

\newbox{\orcid}\sbox{\orcid}{\includegraphics[scale=0.06]{orcid.pdf}} 
\author[1]{%
	\href{https://orcid.org/0000-0000-0000-0000}{\usebox{\orcid}\hspace{1mm}David S.~Hippocampus\thanks{\texttt{hippo@cs.cranberry-lemon.edu}}}%
}
\author[1,2]{%
	\href{https://orcid.org/0000-0000-0000-0000}{\usebox{\orcid}\hspace{1mm}Elias D.~Striatum\thanks{\texttt{stariate@ee.mount-sheikh.edu}}}%
}
\affil[1]{Department of Computer Science, Cranberry-Lemon University, Pittsburgh, PA 15213}
\affil[2]{Department of Electrical Engineering, Mount-Sheikh University, Santa Narimana, Levand}
\fi

\renewcommand{\shorttitle}{\textit{5G NR PRACH Detection with Convolutional Neural Networks (CNN)}}

\hypersetup{
pdftitle={A template for the arxiv style},
pdfsubject={q-bio.NC, q-bio.QM},
pdfauthor={David S.~Hippocampus, Elias D.~Striatum},
pdfkeywords={First keyword, Second keyword, More},
}

\begin{document}
\twocolumn[
\maketitle

\begin{abstract}
In this paper, we present a novel approach to interference detection in 5G New Radio (5G-NR) networks using Convolutional Neural Networks (CNN). Interference in 5G networks challenges high-quality service due to dense user equipment deployment and increased wireless environment complexity. Our CNN-based model is designed to detect Physical Random Access Channel (PRACH) sequences amidst various interference scenarios, leveraging the spatial and temporal characteristics of PRACH signals to enhance detection accuracy and robustness. Comprehensive datasets of simulated PRACH signals under controlled interference conditions were generated to train and validate the model. Experimental results show that our CNN-based approach outperforms traditional PRACH detection methods in accuracy, precision, recall and F1-score. This study demonstrates the potential of AI/ML techniques in advancing interference management in 5G networks, providing a foundation for future research and practical applications in optimizing network performance and reliability.
\end{abstract}

\keywords{5G-NR \and PRACH \and CNN \and  Interference Detection}
\vspace*{+0.5cm}
]

\section{Introduction}
\label{sec:1}

The rapid advancement of wireless communication technologies has led to the development and deployment of 5G New Radio (NR), which aims to meet the ever-increasing demand for high-speed and reliable connectivity. One of the critical components in 5G NR is the Physical Random Access Channel (PRACH), which facilitates the initial access procedure for User Equipment (UE) to establish a connection with the base station (gNB).

PRACH plays a vital role in enabling UEs to identify themselves and initiate communication with the gNB by transmitting a preamble sequence. This process is essential for maintaining seamless connectivity and efficient network operation. However, the presence of cell-interference, which occurs when multiple UEs attempt to access the network simultaneously, poses significant challenges to the accurate detection of PRACH signals.

Traditional methods for PRACH detection rely on correlation-based techniques, which can be susceptible to interference and noise, leading to false alarms and missed detections. To address these challenges, this paper proposes a novel approach utilizing Convolutional Neural Networks (CNNs) to enhance the detection performance of PRACH under cell-interference conditions.

CNNs have demonstrated remarkable success in various domains, particularly in image and signal processing, due to their ability to capture complex patterns and spatial dependencies. By leveraging CNNs for PRACH detection, this study aims to improve the robustness and accuracy of the detection process, thereby reducing the initial access latency and enhancing overall network performance.

The remainder of this paper is organized as follows: Section 2 provides a comprehensive background on 5G NR PRACH and related work. Section 3 outlines the proposed CNN-based PRACH detection model. Section 4 describes the dataset generation and preparation methods. Section 5 details the training and testing methodology. Section 6 presents the results and analysis. Finally, Section 7 concludes the paper and discusses future research directions.

\section{Background and Related Work}
\label{sec:2}
	
The 5G New Radio (NR) introduces significant advancements in wireless communication, with the Physical Random Access Channel (PRACH) playing an important role in enabling the initial attachment of User Equipment (UE) to a Base Station (gNB) \cite{3gpp2018}. The PRACH allows UEs to send a preamble to the gNB, facilitating the identification and synchronization processes necessary for subsequent communication. This section provides an overview of the 5G NR PRACH procedures, the challenges posed by cell interference, and the existing approaches to PRACH detection, emphasizing the transition from conventional methods to AI/ML-based solutions.

\subsection{5G Random Access Procedures}
The 5G NR specifications support two types of Random Access (RA) procedures: contention-based and contention-free \cite{3gpp2018}. In the contention-based RA (CBRA) procedure, UEs randomly select a preamble from a pool, while in the contention-free RA (CFRA) procedure, the gNB assigns specific preambles to UEs. The CBRA procedure involves a four-step process, starting with the transmission of a randomly selected preamble by the UE on the PRACH (Msg 1), followed by the gNB's response with a Random Access Response (RAR, Msg 2). This is succeeded by scheduled uplink and downlink transmissions (Msgs 3 and 4), completing the RA process \cite{ts38113}.

With this information, the UE can transmit the PRACH preamble using the  resources indicated by the Next-Generation NodeB (gNB) during the transmission of SIB2. The transmission of the PRACH is associated with the RA-RNTI \textsl{(Random Access RNTI)}.

\begin{figure*}[h!]
	\centering
\includegraphics[page = 1,clip, trim=0.0cm 0.0cm 0.0cm 0.0cm, width=0.75\textwidth]{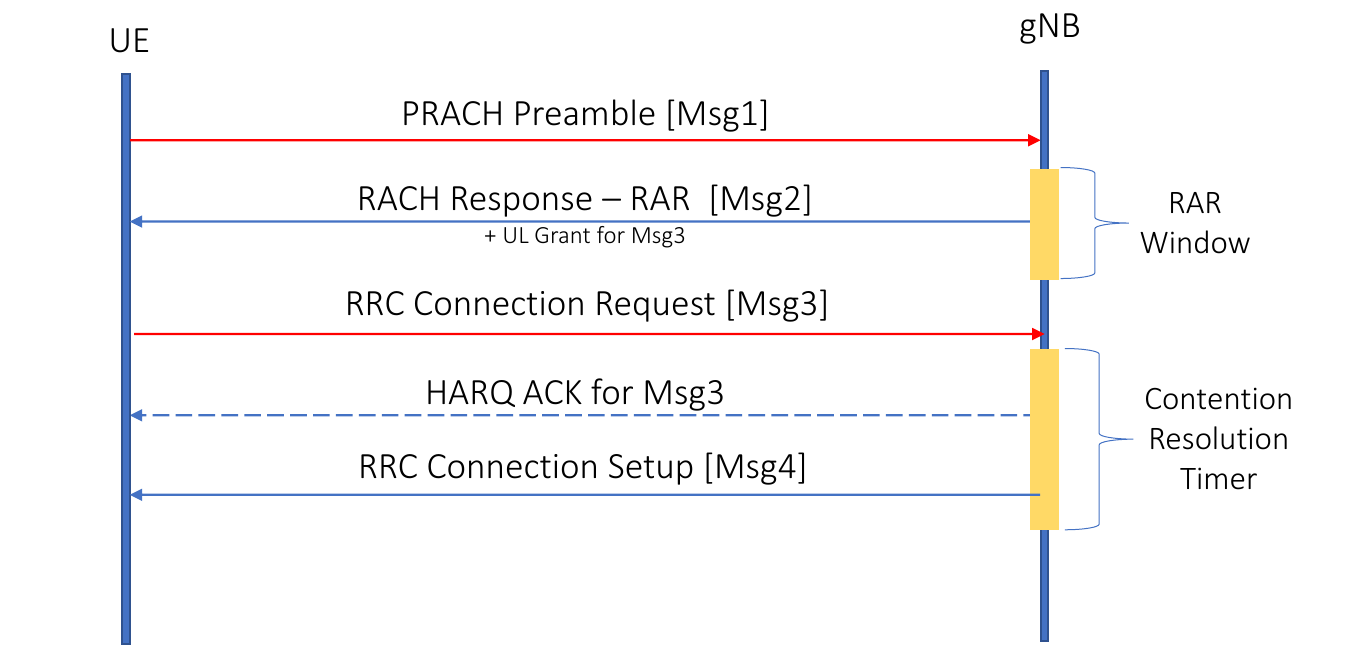}
	\caption{NR-PRACH Procedure.}
	\label{Fig:01}
\end{figure*}

When an UE enters a new cell, it has no prior knowledge of the gNB. After identifying the optimal SSB \textsl{(Synchronization Signal  Block)}  through downlink synchronization, the UE transmits the PRACH containing its information, based on the best time index of the SSB. Fig.\ref{Fig:01} illustrates the interactions between the UE and the gNB during the initial access procedure \cite{AMukherjee2019}.

Random access process is described in Fig.\ref{Fig:01}. The UE selects randomly a preamble from a list of parameters broadcasted through the SIB2 and transmits it in the PRACH with an initial power result of a basic downlink pathloss estimation. If there is no answer from the gNB, the UE makes a retry with higher power level.


\subsection{PRACH Sequence Generation}
\begin{figure}[h!]
	\centering
\includegraphics[page = 1,clip, trim=0.0cm 0.0cm 0.0cm 0.0cm, width=0.50\textwidth]{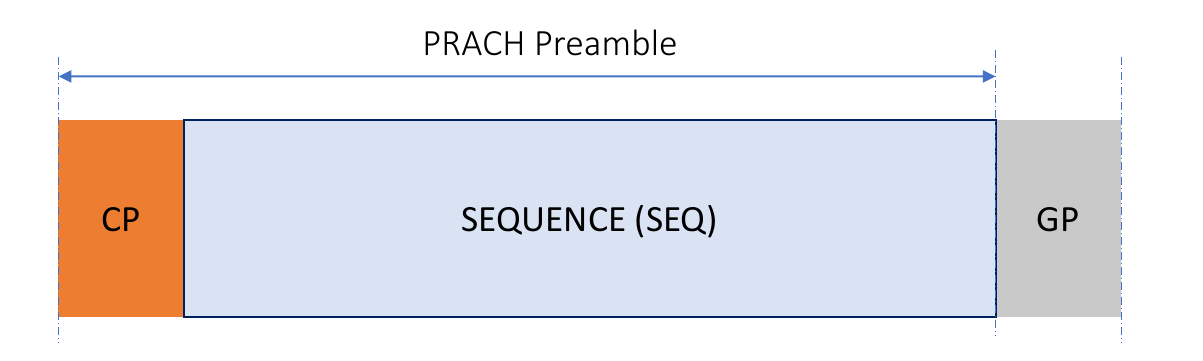}
	\caption{NR-PRACH structure.}
	\label{Fig:02}
\end{figure}

PRACH sequence generation begins with the selection of a base sequence, typically a Zadoff-Chu sequence known for its Constant Amplitude Zero Auto-Correlation (CAZAC) properties \cite{zadoff1963}. The sequence which is which an OFDM symbol, built with a Cyclic Prefix (CP) undergoes a cyclic shift to embed the Random Access Preamble ID (RAPID). The gNB decodes the received sequence by correlating it with the known base sequence, estimating both the RAPID and the Timing Advance (TA), which indicates the propagation delay induced by the UE's distance from the gNB \cite{2401.12803v1}. Refer to Fig.\ref{Fig:02} for the NR-PRACH structure.

The preamble sequence length is set to a prime number of 839, there are 838 sequences with optimal cross‐correlation properties. The ${{\rm{u}}^{{\rm{th}}}}$ $\left({{\rm{0}} \le {\rm{u}} \le {\rm{837}}} \right)$ root Zadoff‐Chu sequence is defined by (Nzc is the length of the Zadoff‐Chu sequence):
\textcolor{myColor}{
\begin{equation}
{x_u}\left( n \right) = {e^{ - j\frac{{\pi un\left( {n + 1} \right)}}{{{N_{ZC}}}}}},0 \le n \le {N_{ZC}} - 1
\label{Eq:01}
\end{equation} 
}

From the ${{\rm{u}}^{{\rm{th}}}}$root ZC sequence, random access preambles with Zero Correlation Zones \textsl{(ZCZ)} of length ${N_{ZC}} - 1$ are defined by Cyclic Shifts \textsl{(CS)} according to \cite{3GPPTS36211V890}:

\textcolor{myColor}{
\begin{equation}
{x_{u,v}}\left( n \right) = {x_u}\left( {\left( {n + {C_v}} \right)\bmod {N_{ZC}}} \right),
\label{Eq:02}
\end{equation}  
}
where ${{C_v}}$ is the cyclic shift, and ${{N_{ZC}}}$ is the cyclic shift offset. This paper adopts preamble format 0 in the 5G-NR, which generates from a 839 point ZC sequence which is  specifically designed for contention-based access, where multiple UEs may attempt to access the network simultaneously.

\subsection{Conventional Approaches for PRACH Detection}
Traditional PRACH receivers rely on correlation-based techniques to detect the RAPID and estimate the TA. These methods involve correlating the received signal with all possible base sequences, identifying the sequence that maximizes the correlation value. The position of the correlation peak indicates the RAPID and TA. However, these methods require setting thresholds to distinguish between correct and incorrect detections, which can be challenging under varying signal conditions and lead to false alarms or missed detections \cite{pham2019, kamata2021}. Additionally, cell interference and multipath effects further complicate the accurate detection of PRACH signals \cite{2401.12803v1}.

\begin{figure*}[htbp]
    \centering
    \includegraphics[page = 1, clip, trim=0.0cm 0.0cm 0.0cm 0.0cm, width=0.85\textwidth]{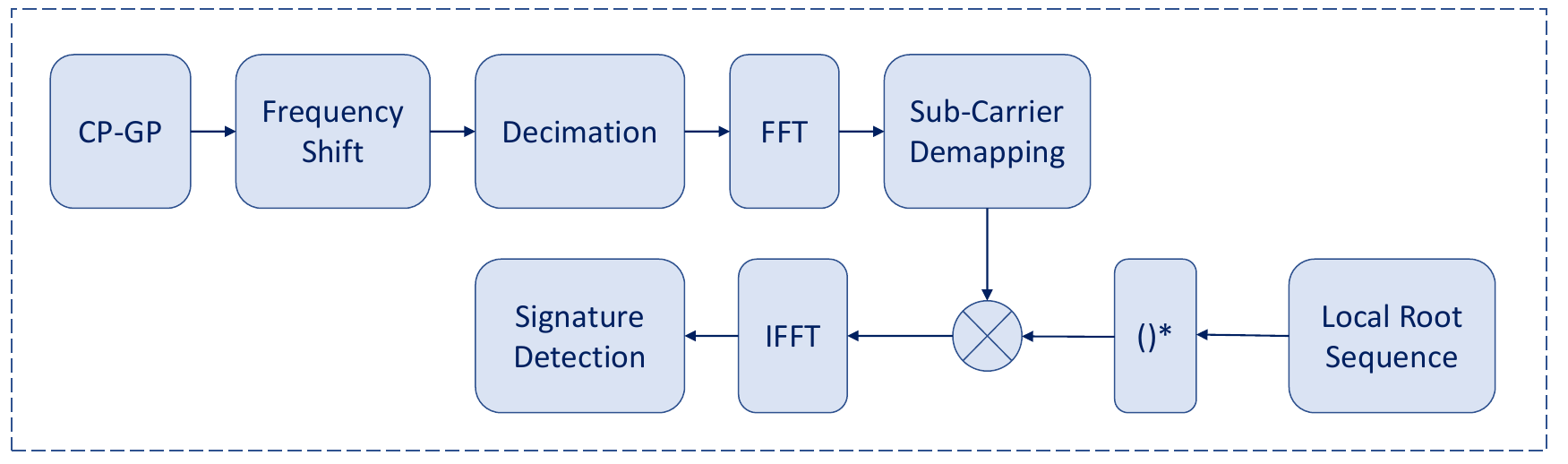}
    \caption{Functional diagram of the 5G-NR PRACH receiver.}
    \label{Fig:03}
\end{figure*}

Figure \ref{Fig:03} illustrates the block diagram of the NR-PRACH receiver in 5G-NR networks. It is impotant to understand the PRACH signal processing flow in 5G-NR networks. Each module in the diagram plays a critical role in extracting and identifying random access signals, and interference at any stage can compromise detection quality. By integrating CNN techniques, the goal is to enhance the robustness and accuracy of these processing steps against interference, thereby optimizing the overall performance of the PRACH receiver.

\subsection{AI Applications in Wireless Communications}


The Table \ref{Table:RelatedWork}  categorizes the relevant literature into four distinct and meaningful categories. These categories—Interference Detection, AI/ML for Wireless Communications, Convolutional Neural Networks (CNN) and 5G and PRACH—cover the broad spectrum of research areas pertinent to the topic. 

\begin{table*}[htbp]
\centering
\resizebox{0.85\textwidth}{!}{
\begin{tabular}{|p{4cm}|p{2cm}|p{7cm}|}
\hline
Category & Reference & Description \\ \hline

Interference Detection & \cite{cheraghinia2024explainable}, \cite{raj2024dynamic}, \cite{wang2023study}, \cite{massimi2023artificial}, \cite{robinson2023narrowband}, \cite{he2022novel}, \cite{ali2020adversarial}, \cite{zilz2019optimal} & Techniques for detecting wireless interference using AI and deep learning methods. \\ \hline

AI/ML for Wireless Communications & \cite{madasamy2023novel}, \cite{akila2023forecasting}, \cite{li2023intelligent}, \cite{xia2022deep}, \cite{hussain2022jamming}, \cite{xu2022real}, \cite{swinney2021rf}, \cite{elgebali2021multi} & Applications of AI/ML in improving wireless communications, from signal detection to network optimization. \\ \hline

Convolutional Neural Networks (CNN) & \cite{sagar2023wireless}, \cite{chen2023reservoir}, \cite{zhang2022deep} & Use of CNNs in wireless communications for signal processing tasks. \\ \hline

5G and PRACH & \cite{noori2023detection}, \cite{zhang2022deep} & Focus on 5G technology and PRACH signal detection challenges. \\ \hline

\end{tabular}
}
\caption{Interference Detection and AI/ML Applications in Wireless Communications}
\label{Table:RelatedWork}
\end{table*}


\subsection{AI/ML Approaches for PRACH Detection}
Recent advancements in AI and machine learning have introduced new methods for PRACH detection. These approaches leverage neural networks to classify RAPID and TA directly from frequency domain data, bypassing the need for correlation \cite{jang2021, zehra2022}. AI/ML models, such as Convolutional Neural Networks (CNNs), have shown improved performance in detecting PRACH signals, particularly in challenging environments with low Signal-to-Noise Ratio (SNR) and high interference \cite{Fang2022}.

Unlike traditional methods, AI/ML-based receivers can operate in parallel, with separate models for RAPID detection and TA estimation. This parallel processing reduces the error propagation between sequential detection steps, enhancing overall detection accuracy \cite{2401.12803v1}. Furthermore, AI/ML models can generalize across different channel conditions and hardware impairments, making them robust solutions for real-world deployments \cite{2401.12803v1}.

\subsection{AI/ML Techniques in Mobile Networks}
AI/ML techniques revolutionize the management and optimization of mobile networks by providing advanced solutions to enhance performance, reliability, and efficiency in increasingly complex environments. Table \ref{tab:01} summarizes various studies and applications of AI/ML techniques in mobile networks. It illustrates the multiple contexts and models used to improve performance and manage interference in 5G networks and beyond.

\begin{table*}[htbp]
\centering
\resizebox{0.85\textwidth}{!}{
\begin{tabular}{|p{1.5cm}|p{5cm}|c|p{5cm}|}
\hline
\textbf{Reference} & \textbf{Context} & \textbf{Model AI/ML}  & \textbf{Application} \\ \hline
Dahal et al.\cite{Dahal2022} & DRL for interference management & DRL & Millimeter-wave networks \\ \hline
Fang et al.\cite{Fang2022} & PRACH detection via FNN & FNN & PRACH detection for 5G \\ \hline
Dahal et al.\cite{Dahal2023} & DRL for interference mitigation & DRL & Multi-cellular interference management \\ \hline
Zhang et al.\cite{Zhang2018} & Deep Learning in Networking & DL & Mobile and wireless networks \\ \hline
ITU \cite{Itu2020} & ITU AI/ML 5G Challenge & AI/ML & 5G networks \\ \hline
Guel et al. \cite{Guel2023} & Optimization of PRACH detection performance in 5G-NR networks & - & Matlab simulation, 5G-NR interference \\ \hline
Misc \cite{Misc2021} & Radio access procedure and 5G-NR characteristics & - & 5G-NR \\ \hline
Intel \cite{intelAI5G} & AI for 5G network optimization & AI/ML & 5G network optimization \\ \hline
Launay \cite{Launay2021} & Description of NG-RAN and 5G-NR & - & 5G deployment \\ \hline
Cox \cite{Cox2021} & Comprehensive introduction to 5G & - & 5G introduction \\ \hline
Moltchanov et al. \cite{Moltchanov2022} & Mathematical modeling of millimeter-wave and terahertz 5G/6G cellular systems & - & Mathematical modeling \\ \hline
Yaro et al. \cite{Yaro2022} & 5G NR robustness study & - & Telecommunications \\ \hline
Boumaza et al. \cite{Boumaza2020} & Deep learning evaluation in 5G & DL & 5G-NR \\ \hline
Xiong et al. \cite{Xiong2018} & Design of random access preambles for 5G-NR & - & 5G-NR \\ \hline
Zhang et al. \cite{Zhang2019} & Deep Learning for Mobile Networking & DL & Mobile Networking \\ \hline
Zhang et al. \cite{Zhang2021} & Online VNF Deployment for 5G Network Slice & - & 5G Network Slice \\ \hline
Hassan et al. \cite{Hassan2021} & AI Techniques for 5G Mobile Networks & ML/DL & 5G Mobile Networks \\ \hline
\end{tabular}
}
\caption{AI/ML Applications in mobile networks}
\label{tab:01}
\end{table*}

\subsubsection{Service Optimization}
In the domain of service optimization, several techniques are deployed to ensure the proper functioning of equipment and the network in real-time.


\begin{itemize}
    \item \textcolor{myColor}{Predictive Maintenance}: Use of supervised learning to identify and anticipate failures based on historical malfunction data and user experience. This allows for early recognition of malfunction symptoms and detection of new types of problems \cite{raj2024dynamic, wang2023study}.
    \item \textcolor{myColor}{Real-time Equipment and Network Optimization}: AI can optimize energy consumption by putting unused equipment to sleep or dynamically allocating more resources to active equipment \cite{madasamy2023novel}.
    \item \textcolor{myColor}{Traffic Classification}: Machine learning-based traffic classification algorithms identify different types of data flows based on their characteristics. However, classification errors can affect net neutrality \cite{massimi2023artificial, chen2023reservoir}.
    \item \textcolor{myColor}{Quality of Service (QoS) Optimization}: AI is used to manage many parameters influencing service quality, which would be impossible to do manually due to the complexity of dependencies between parameters \cite{zhang2022deep, xu2022real}.
\end{itemize}

\subsubsection{Radio Planning}
Radio planning in 5G mobile networks integrates advanced strategies such as beamforming and machine learning for optimized cell deployment, addressing the challenges of the next generation of mobile communications:


\begin{itemize}
    \item \textcolor{myColor}{Beamforming}: Using multiple antennas to create constructive and destructive interference patterns to direct the radio beam. AI helps to parameterize these signals to optimize the direction of transmission and minimize interference \cite{hussain2022jamming, he2022novel}.
    \item \textcolor{myColor}{Optimized Cell Deployment}: AI optimizes the location of small cells to maximize service quality while minimizing deployment costs \cite{swinney2021rf, zilz2019optimal}.
\end{itemize}


\section{Proposed CNN-based PRACH Interference Detection Model}
\label{sec:3}



In this section, we present the framework adopted to address the issue of interference management in 5G-NR networks. Our approach focuses on the use of innovative techniques based on Artificial Intelligence (AI) and Machine Learning (ML) to classify PRACH signals \cite{Xiong2018,Fang2022,pham2019} affected by interference. We will describe the system architecture and discuss the AI/ML models chosen for interference detection/classification. We will also explain the performance evaluation methods used to assess the effectiveness of our approach.

\subsection{Methodological Approach}
Our approach relies on the use of Convolutional Neural Networks (CNN) \cite{kattenborn2021review} for detecting the presence of cellular interference in PRACH signal detection. We exploit the spatio-temporal characteristics of PRACH signals \cite{Xiong2018,Fang2022,pham2019}, as well as channel information, to distinguish between unperturbed signals and signals disturbed by interference. The Figure \ref{fig:simulation_workflow} shows the   multi-step approach which allows us to efficiently detect and characterize interference in PRACH signal detection.

\tikzstyle{block} = [rectangle, draw, fill=gray!10, 
    text width=10em, text centered, rounded corners, minimum height=3em]
\tikzstyle{line} = [draw, ->]
\tikzstyle{startstop} = [ellipse, draw, fill=gray!10, 
    text width=8em, text centered, minimum height=2em]

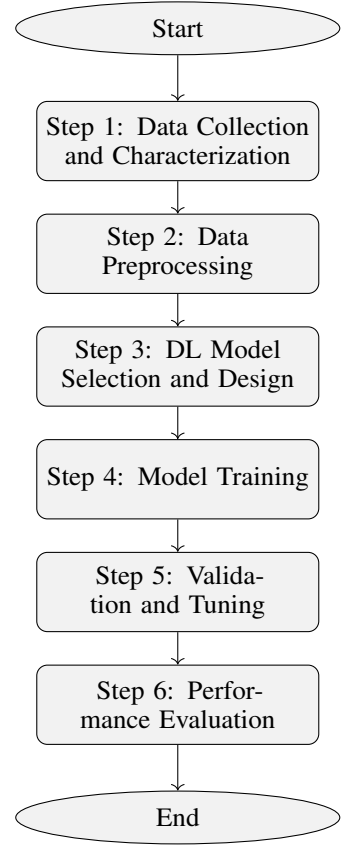
\begin{figure}[h]
\centering
\begin{tikzpicture}[node distance=1.5cm, auto]

\node [startstop] (start) {Start};
\node [block, below of=start] (step1) {Step 1: Data Collection and Characterization};
\node [block, below of=step1] (step2) {Step 2: Data Preprocessing};
\node [block, below of=step2] (step3) {Step 3: DL Model Selection and Design};
\node [block, below of=step3] (step4) {Step 4: Model Training};
\node [block, below of=step4] (step5) {Step 5: Validation and Tuning};
\node [block, below of=step5] (step6) {Step 6: Performance Evaluation};
\node [startstop, below of=step6] (end) {End};

\path [line] (start) -- (step1);
\path [line] (step1) -- (step2);
\path [line] (step2) -- (step3);
\path [line] (step3) -- (step4);
\path [line] (step4) -- (step5);
\path [line] (step5) -- (step6);
\path [line] (step6) -- (end);

\end{tikzpicture}
    \caption{ML model for interference detection/classification.}
    \label{fig:simulation_workflow}
\end{figure}

\begin{itemize}
    \item \textcolor{myColor}{(i) Data Collection and Characterization:} In this initial step, we collect PRACH signal data from simulated scenarios representative of real-world 5G-NR network conditions. These data are characterized by parameters such as signal power level, Signal-to-Noise Ratio (SNR), and present interferences.
    \item  textcolor{myColor}{(ii) Data Preprocessing:} The collected data are preprocessed to make them suitable for input to Deep Learning (DL) models. This includes normalization, filtering, dimensionality reduction, etc. The goal is to clean the data and extract relevant features for classification.
    \item \textcolor{myColor}{(iii) DL Model Selection and Design:} We select and design an appropriate Machine Learning (ML) model for PRACH signal classification. In our case, we opted for a Convolutional Neural Network (CNN) \cite{kattenborn2021review} due to its ability to capture spatial and temporal features of the data.
    \item \textcolor{myColor}{(iv) Model Training:} We use the preprocessed data to train the selected AI/ML model. During training, the model learns to recognize patterns and relationships between input data and output labels, indicating the presence or absence of interference.
    \item \textcolor{myColor}{(v) Validation and Tuning:} After training, we evaluate the model's performance on a separate validation dataset. This allows us to verify if the model generalizes well to unseen data and identify potential overfitting issues. 
    \item \textcolor{myColor}{(vi) Performance Evaluation:} Finally, we evaluate the performance of the final model on an independent test dataset. We use metrics such as precision, recall, and F1-score to measure classification effectiveness and compare our approach to other existing methods.
\end{itemize}

\subsection{AI/ML Model Selection for Interference Management}
We chose to use Convolutional Neural Networks (CNN) due to their ability to capture the spatial and temporal characteristics of PRACH signals. CNNs are particularly well-suited for image and sequence classification, making them a natural choice for our signal classification problem. The CNN model comprises:
\begin{itemize}
    \item Convolutional layers to extract local features from PRACH signals.
    \item Pooling layers to reduce dimensionality and prevent overfitting.
    \item Fully connected layers for final signal classification.
    \item Activation functions (such as ReLU) to capture non-linear relationships.
\end{itemize}

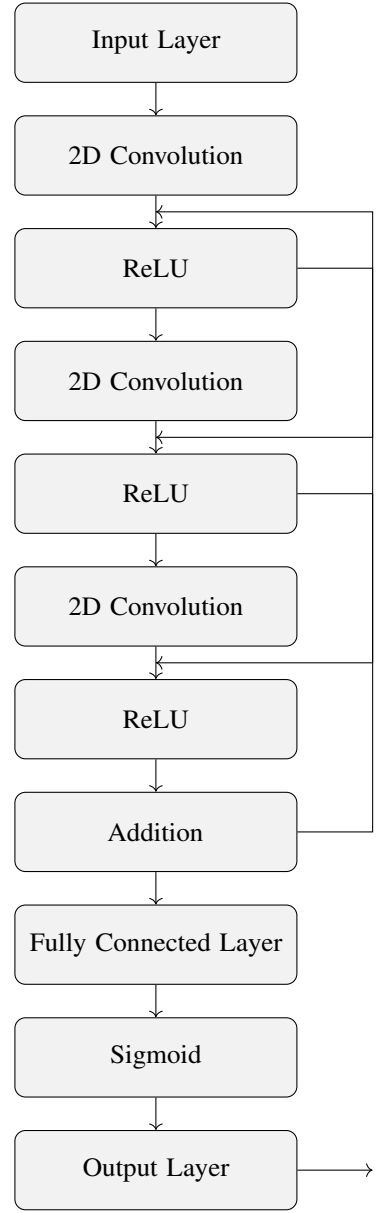
\begin{figure}[htbp]
    \centering
    \begin{tikzpicture}[node distance = 1.5cm, auto]
        \node [block] (input) {Input Layer};
        \node [block, below of=input] (conv1) {2D Convolution};
        \node [block, below of=conv1] (relu1) {ReLU};
        \node [block, below of=relu1] (conv2) {2D Convolution};
        \node [block, below of=conv2] (relu2) {ReLU};
        \node [block, below of=relu2] (conv3) {2D Convolution};
        \node [block, below of=conv3] (relu3) {ReLU};
        \node [block, below of=relu3] (add) {Addition};
        \node [block, below of=add] (fc) {Fully Connected Layer};
        \node [block, below of=fc] (sigmoid) {Sigmoid};
        \node [block, below of=sigmoid] (classoutput) {Output Layer};

        \path [line] (input) -- (conv1);
        \path [line] (conv1) -- (relu1);
        \path [line] (relu1) -- (conv2);
        \path [line] (conv2) -- (relu2);
        \path [line] (relu2) -- (conv3);
        \path [line] (conv3) -- (relu3);
        \path [line] (relu3) -- (add);
        \path [line] (add) -- (fc);
        \path [line] (fc) -- (sigmoid);
        \path [line] (sigmoid) -- (classoutput);

        \draw[->] (add.east) -- ++(1,0) |- node [near end] {} ($(conv1)!0.5!(relu1)$);
        \draw[->] (relu1.east) -- ++(1,0) |- node [near end] {} ($(conv2)!0.5!(relu2)$);
        \draw[->] (relu2.east) -- ++(1,0) |- node [near end] {} ($(conv3)!0.5!(relu3)$);
        \draw[->] (classoutput.east) -- ++(1,0);
    \end{tikzpicture}
    \caption{Different layers Diagram of the CNN model.}
    \label{fig:CNN_diagram}
\end{figure}

\begin{table*}[h!]
    \centering
		\resizebox{0.85\textwidth}{!}{
    \begin{tabular}{|c|c|c|p{3cm}|}
    \hline
    {Layer} & {Type} & {Parameters} & {Remarks} \\ \hline
    \multirow{2}{*}{{Input}} & \multirow{2}{*}{ImageInputLayer} & \multicolumn{2}{p{6cm}|}{Dimensions: [length(SNRdBInterf), length(SNRdB), chcfg.NRxAnts*1938]} \\ \cline{3-4} 
     &  & Name: input & SplitComplexInputs: true \\ \hline
    \textcolor{myColor}{Conv1} & Convolution2DLayer & Filter: $5 \times 5$, Outputs: 20 & Name: conv\_1 \\ \hline
    \textcolor{myColor}{ReLU1} & ReLULayer &  & Name: relu\_1 \\ \hline
    \textcolor{myColor}{Conv2} & Convolution2DLayer & Filter: $3 \times 3$, Outputs: 20, Padding: 1 & Name: conv\_2 \\ \hline
    \textcolor{myColor}{ReLU2} & ReLULayer &  & Name: relu\_2 \\ \hline
    \textcolor{myColor}{Conv3} & Convolution2DLayer & Filter: $3 \times 3$, Outputs: 20, Padding: 1 & Name: conv\_3 \\ \hline
    \textcolor{myColor}{ReLU3} & ReLULayer &  & Name: relu\_3 \\ \hline
    \textcolor{myColor}{Add} & AdditionLayer & Inputs: 2 & Name: add \\ \hline
    \textcolor{myColor}{FC} & FullyConnectedLayer & Neurons: 2 & Name: fc \\ \hline
    \textcolor{myColor}{Sigmoid} & SigmoidLayer &  & Name: sigmoid \\ \hline
    \textcolor{myColor}{Classification} & ClassificationLayer &  & Name: classoutput \\ \hline
    \end{tabular}
		}
    \caption{CNN model layers Configuration}
    \label{tab:CNN_layers}
\end{table*}

\subsection{Performance Evaluation Methods}
To evaluate the performance of our approach, we use standard metrics such as accuracy, recall, and F1-score. These metrics allow us to quantify classification effectiveness and compare different approaches. We also use cross-validation and data partitioning techniques to assess model generalization and prevent overfitting.

\section{Training and Testing Methodology}
\label{sec:5}

This section details the methodological approach adopted to address the issue of interference management in 5G-NR networks. We emphasize the design and implementation of innovative techniques based on Artificial Intelligence (AI) and Machine Learning (ML).

\subsection{Proposed Approach}
Our approach focuses on using Convolutional Neural Networks (CNN) for detecting cellular interferences in PRACH (Physical Random Access Channel) signal detection. We exploit the spatio-temporal characteristics of PRACH signals and channel information to distinguish between undisturbed and disturbed signals. This approach involves several successive steps, allowing us to effectively detect and characterize interferences in PRACH signal detection.

\subsection{System Architecture Presentation}
Figure \ref{Fig:03} illustrates the block diagram of the NR-PRACH receiver in 5G-NR networks. It is important to understand the PRACH signal processing flow in 5G-NR networks. Each module in the diagram plays an important role in extracting and identifying random access signals, and any interference at any stage can compromise detection quality. By integrating Convolutional Neural Networks (CNN) techniques, the goal is to improve the robustness and accuracy of these processing stages against interferences, thus optimizing the overall performance of the PRACH receiver.

The different signal processing steps from initial reception to signature detection are detailed as follows:
\begin{itemize}
    \item \textsl{\textcolor{myColor}{(i) CP-GP (Cyclic Prefix - Guard Period):}} This step removes the cyclic prefix and guard period from the received signal, preparing it for frequency processing.
    \item \textsl{\textcolor{myColor}{(ii) Frequency Shift:}} The signal is then subjected to a frequency shift to correctly align the received signal frequencies with those of the receiver.
    \item \textsl{\textcolor{myColor}{(iii) Decimation:}} This step reduces the signal's sampling rate, decreasing the amount of data to be processed while retaining essential information.
    \item \textsl{\textcolor{myColor}{(iv) FFT (Fast Fourier Transform):}} FFT is applied to convert the signal from the time domain to the frequency domain, facilitating sub-carrier analysis and processing.
    \item \textsl{\textcolor{myColor}{(v) Sub-Carrier Demapping:}} This step demaps the sub-carriers to extract specific data from the frequency sub-channels used in PRACH transmission.
    \item \textsl{\textcolor{myColor}{(vi) Local Root Sequence:}} A local root sequence is used for correlation with the received signal to facilitate PRACH signature detection.
    \item \textsl{\textcolor{myColor}{(vii) Conjugation:}} Complex conjugation is applied to the signal to prepare the data for inverse Fourier transformation.
    \item \textsl{\textcolor{myColor}{(viii) IFFT (Inverse Fast Fourier Transform):}} IFFT is applied to convert the signal back from the frequency domain to the time domain.
    \item \textsl{\textcolor{myColor}{(ix) Signature Detection:}} The final step involves correlating the processed signal with the local root sequence to detect the PRACH signature, determining the presence and identity of the access terminal.
\end{itemize}

\section{Results and Analysis}
\label{sec:6}

	%

In this section, we present the results obtained from the PRACH detection simulations in the presence of interference. We also describe the interpretation of these results based on different simulation parameters.

\subsection{Description of Simulated Data}
This section details the configurations and parameters used to generate the PRACH simulation data, including UE parameters, interference configuration and the signal-to-noise ratio (SNR) scenarios studied.

\begin{table*}[htbp]
\centering
\caption{Configurations for UE, PRACH and Interference}
		\resizebox{0.85\textwidth}{!}{
\begin{tabular}{|p{4cm}|p{4cm}|p{5cm}|}
\hline
Configuration & Parameter & Value/Description \\ \hline

UE Configuration & NULRB & 6 (Number of Resource Blocks) \\ \cline{2-3}
 & Duplex Mode & FDD \\ \cline{2-3}
 & Cyclic Prefix UL & Normal \\ \cline{2-3}
 & NTxAnts & 1 (Number of Transmission Antennas) \\ \hline

{PRACH Configuration} & Format & 0 (Low Mobility) \\ \cline{2-3}
 & Sequence Index & 22 \\ \cline{2-3}
 & Cyclic Shift Index & 1 \\ \cline{2-3}
 & High-speed Mode & Disabled \\ \cline{2-3}
 & Frequency Offset & 0 \\ \cline{2-3}
 & Preamble Index & 32 \\ \hline

{Interference Configuration} & PRACH Format & 0 (Low Mobility) \\ \cline{2-3}
 & Sequence Index & 22 \\ \cline{2-3}
 & Cyclic Shift Index & 1 \\ \cline{2-3}
 & High-speed Mode & Disabled \\ \cline{2-3}
 & Frequency Offset & 0 \\ \cline{2-3}
 & Preamble Index & 3 \\ \hline

\end{tabular}
}
\label{table:config_summary}
\end{table*}

The Table \ref{table:config_summary} summarizes the key configuration parameters for  UE, PRACH, and PRACH Interference used in the 5G-NR simulations. It includes essential details such as the number of resource blocks, duplex mode, cyclic prefix, and preamble index, which are crucial for accurately simulating and evaluating PRACH detection under various conditions. This structured overview facilitates understanding and comparison of the configurations applied in the study.

\subsection{Data Processing Methodology}
In this section, we describe the methodology adopted for generating and processing the simulation data used to evaluate the performance of the interference detection model in PRACH signals. We address the key steps in simulating PRACH and interference signal configurations, modeling the propagation channel, and preparing the data for training and validating the neural network model as illustrated in Figure \ref{fig:simulation_workflow}. The process is designed to reproduce realistic and varied test conditions, thus enabling a thorough system evaluation. Each simulation step is configured to explore the impact of different SNR levels and interference power, providing a diverse dataset for algorithm development and verification.

\subsubsection{Data Generation}
PRACH data is generated for different SNR levels and interference power levels relative to the signal. The tested SNR levels are: -18, -15, -12, -9, -6 dB. The tested interference levels are: -30, -27, -21, -15, -12, -9, -6 dB.

The selected SNR values, ranging from -18 dB to -6 dB, allow for evaluating the PRACH detection system's performance under extreme low-signal conditions, which is essential for testing system robustness in challenging reception scenarios. The interference power levels, ranging from -30 dB to -6 dB, are chosen to simulate a realistic range of environments with varying interference intensities to study their impact on detection accuracy under different conditions.

\subsubsection{Propagation Channel Modeling}
The modeling of the propagation channel plays a crucial role in analyzing the performance of PRACH detection techniques in the presence of interference. For our simulations, we adopted a channel model conforming to the 3GPP TS36.104 \cite{3gppLTE104} standard, specifically tailored for large-scale urban mobile communication scenarios.

\paragraph{Channel Model Parameters}
The key parameters used in configuring our channel model are as follows:
\begin{itemize}
    \item \textsl{\textcolor{myColor}{Number of Reception Antennas (NRxAnts)}}: We used two reception antennas to simulate a realistic MIMO environment and to assess the effect of reception diversity on PRACH detection performance.
    \item \textsl{\textcolor{myColor}{Delay Profile}}: The "Extended Typical Urban" (ETU) delay profile was used. This profile is characterized by dense echo and high reverberation, typical of urban environments where signals are reflected by multiple buildings.
    \item \textsl{\textcolor{myColor}{Doppler Frequency}}: The Doppler frequency was set to 70 Hz to simulate the effect of user mobility at moderate speeds, which is relevant for urban scenarios.
    \item \textsl{\textcolor{myColor}{MIMO Correlation}}: The correlation between antennas was defined as low, assuming sufficient separation between antennas to minimize interference between them.
    \item \textsl{\textcolor{myColor}{Sampling Rate}}: The sampling rate was determined by the PRACH configuration, ensuring that the temporal resolution is sufficient to capture fine channel variations due to the relative movement between transmitter and receiver.
\end{itemize}

\paragraph{Channel Model Implementation}
The propagation channel simulation was implemented using MATLAB's fading channel functions \cite{matlab2022}. These functions generate a channel sample that is applied to the transmitted signals to simulate the effects of multipath and relative movement. The Rayleigh fading model\footnote{The Rayleigh fading model is a widely used model to describe the effects of fading (random signal attenuation) in wireless communication systems or mobile networks. It is named after physicist Lord Rayleigh, who contributed to its development in the early 20th century.}, specified by the \texttt{ModelType = 'GMEDS'} option, was used to introduce realistic fading in accordance with the ETU delay profile.

\subsection{Neural Network Architecture}
The convolutional neural network architecture used for PRACH signal detection under heavy interference conditions is described in Figure \ref{fig:CNN_diagram}. The architecture is designed to efficiently extract complex features from the signal data and perform accurate classification.

The network consists of several convolutional layers, activation layers, and pooling layers, followed by a fully connected layer and an output layer for classification. This structure helps capture both local and global patterns in the input data.

The technical details of the layer configurations, such as the number of filters, filter sizes, strides and activation functions are detailed in Figure \ref{fig:CNN_diagram} and are essential for optimizing network performance. These parameters are adjusted based on the results of cross-validation phases during training.

\subsection{Model Training Results}
The results of the model training show the accuracy and validation loss measured over several epochs. The figures below illustrate the evolution of accuracy and loss during training.

\begin{figure*}[htbp]
    \centering
    \includegraphics[page=1, clip, trim=0.0cm 0.0cm 0.0cm 0.0cm, width=0.85\textwidth]{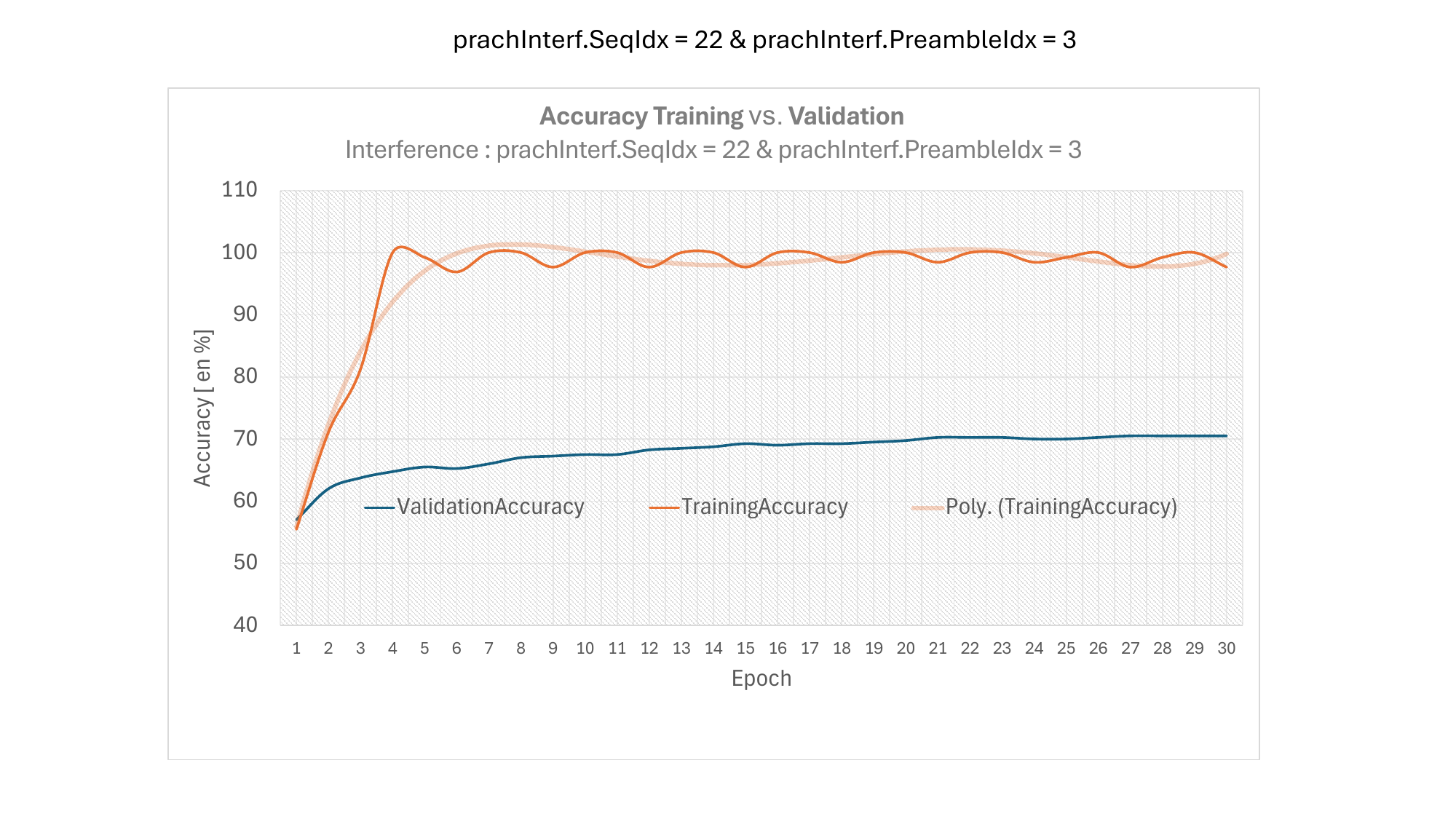}
    \caption{Training vs. Validation Accuracy of the CNN Model}
    \label{fig:Figure2Resultats_1}
\end{figure*}

Figure \ref{fig:Figure2Resultats_1} shows the evolution of the training and validation accuracy of the neural network model over 30 epochs. The training accuracy curve increases rapidly at the beginning and reaches about 95\% after ten epochs, then stabilizes. In contrast, the validation accuracy curve increases more slowly and stabilizes around 75-80\%. This gap between the training and validation curves indicates some overfitting, where the model performs very well on the training data but less so on the validation data. Note that model validation uses a dataset generated with the same values of UE, cell, system parameters, etc.

\begin{figure*}[htbp]
    \centering
    \includegraphics[page=2, clip, trim=0.0cm 0.0cm 0.0cm 0.0cm, width=0.85\textwidth]{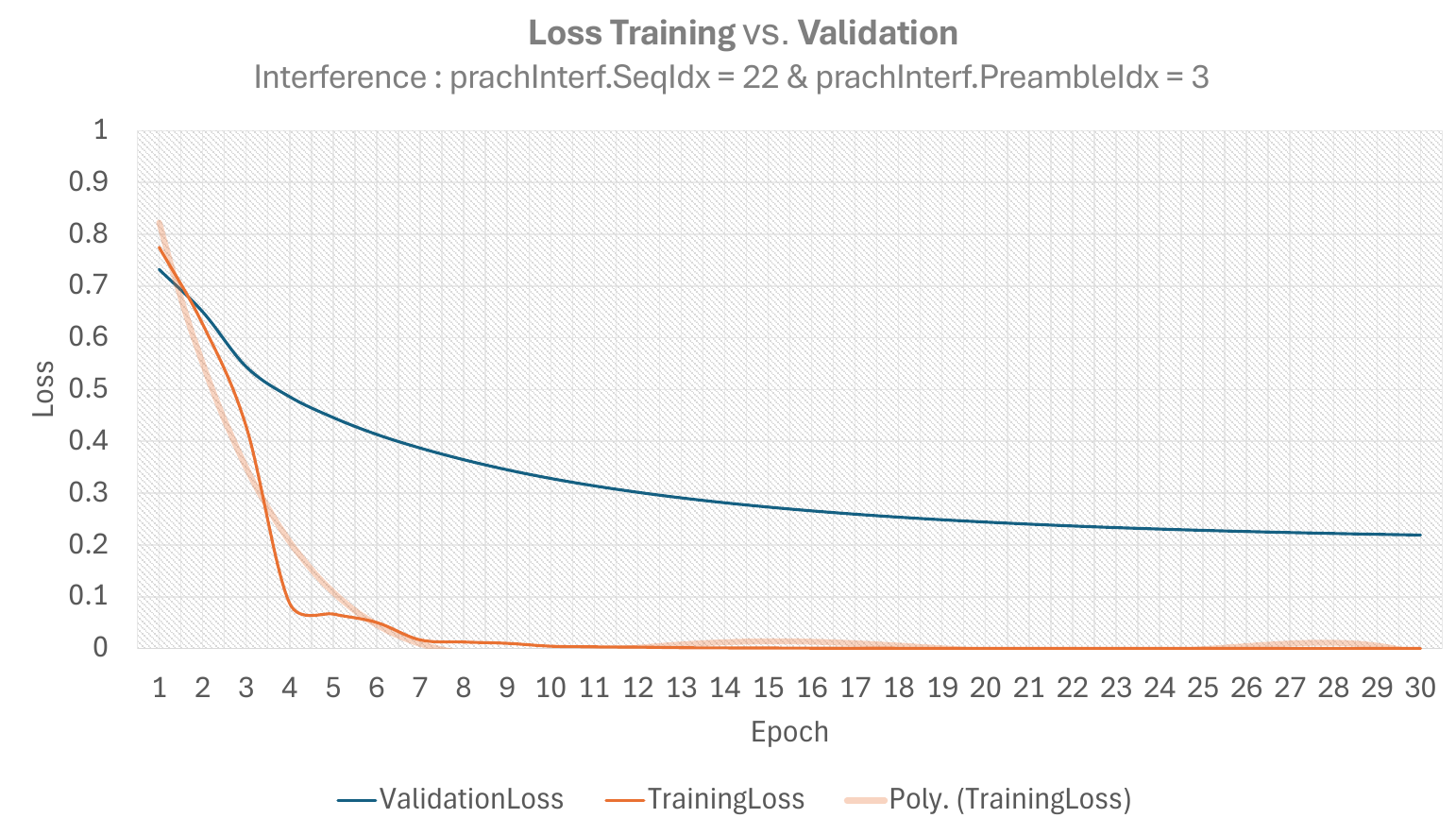}
    \caption{Training vs. Validation Loss of the CNN Model}
    \label{fig:Figure2Resultats_2}
\end{figure*}

Figure \ref{fig:Figure2Resultats_2} presents the training and validation loss curves of our model over 30 epochs. The training loss curve shows a rapid decrease in the first few epochs, reaching an almost negligible value after about five epochs, then stabilizing. In contrast, the validation loss curve decreases more gradually and stabilizes around 0.3 after about ten epochs. Again, this difference between the two curves suggests that the model quickly adapts to the training data but struggles to generalize this knowledge to unseen validation data.

The joint analysis of Figures \ref{fig:Figure2Resultats_1} and \ref{fig:Figure2Resultats_2} reinforces the diagnosis of overfitting. The rapid convergence and high training accuracy values combined with an almost negligible training loss show that the model has learned the specific patterns of the training data. However, the inability to achieve similar performance on the validation data suggests that the model has memorized the training data rather than learning generalizable features.

\begin{figure*}[htbp]
    \centering
    \includegraphics[page=3, clip, trim=0.0cm 0.0cm 0.0cm 0.0cm, width=0.90\textwidth]{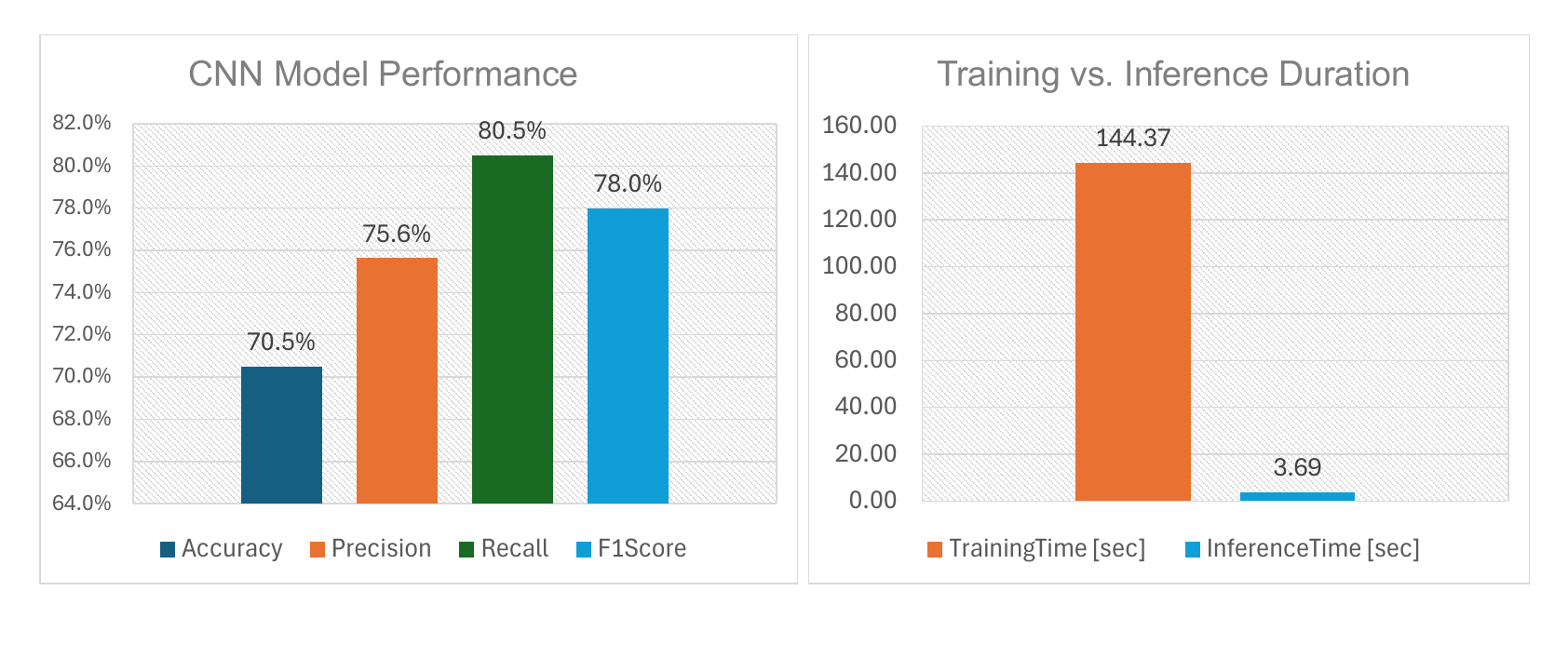}
    \caption{CNN Model Performance (Accuracy, Precision, Recall, F1-Score) and Processing Time}
    \label{fig:Figure2Resultats_3}
\end{figure*}

Figure \ref{fig:Figure2Resultats_3} provides an overview of the performance of our CNN model in terms of PRACH interference classification metrics and processing time (training and inference). An accuracy of 70.5\%, precision of 75.6\%, recall of 80.5\%, and F1 score of 78.0\% suggest that the model has a good ability to correctly identify relevant samples with particularly high recall. This means that the model is effective at detecting true positives (interference). However, the relatively lower accuracy compared to other metrics suggests that the model might still confuse some samples. The F1 score, which combines precision and recall, shows an overall balanced performance at 78.0\%. The relatively long training time (144.37 seconds) reflects the model's complexity and the volume of data used for training. In contrast, the very short inference time (3.69 seconds) indicates that once trained, the model can make predictions quickly and efficiently, which is necessary for real-time applications.

\begin{figure*}[htbp]
    \centering
    \includegraphics[page=4, clip, trim=0.0cm 0.0cm 0.0cm 0.0cm, width=0.85\textwidth]{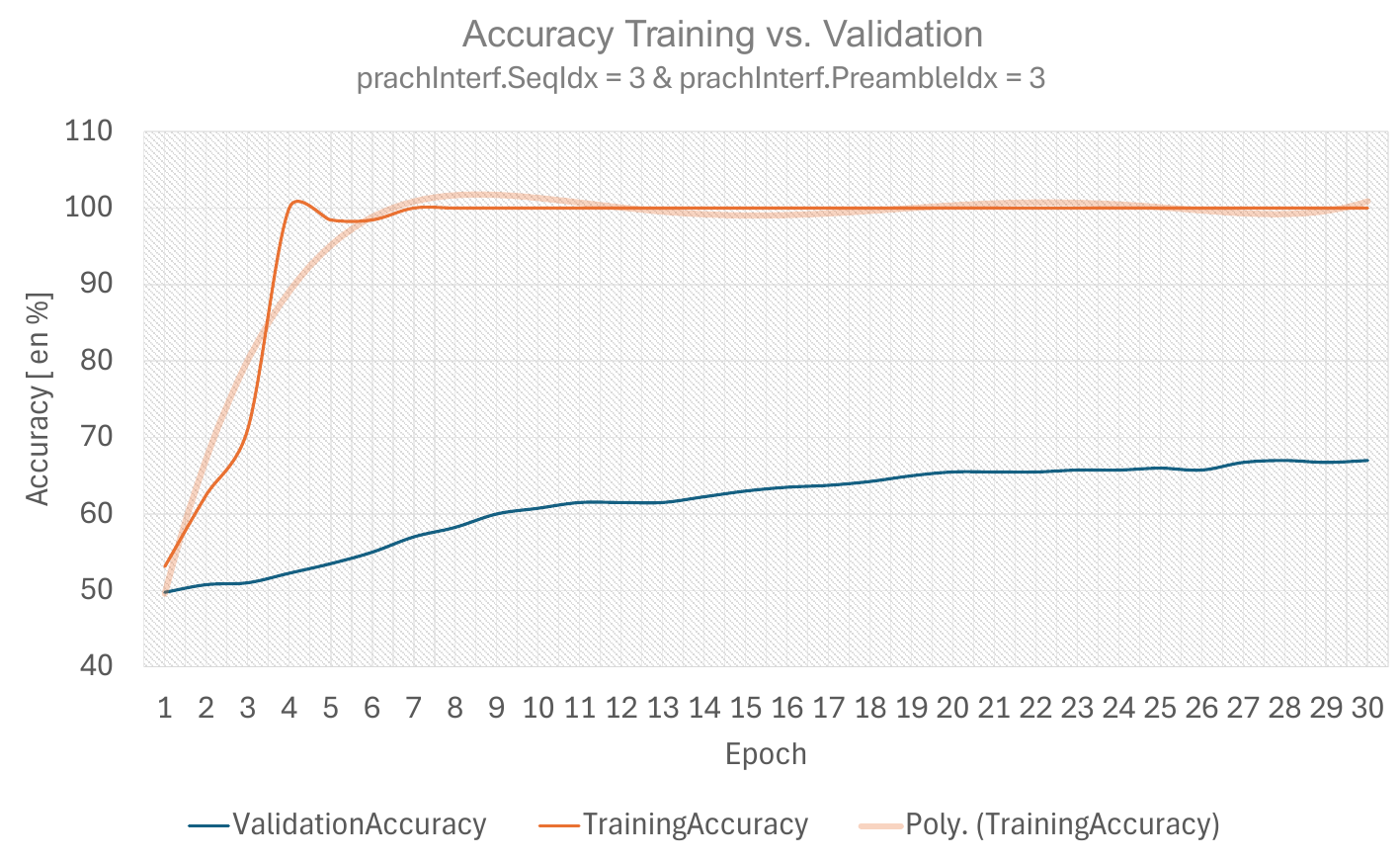}
    \caption{Training vs. Validation Accuracy Parameter Changes Between Training and Inference}
    \label{fig:Figure2Resultats_4}
\end{figure*}

\begin{figure*}[htbp]
    \centering
    \includegraphics[page=5, clip, trim=0.0cm 0.0cm 0.0cm 0.0cm, width=0.85\textwidth]{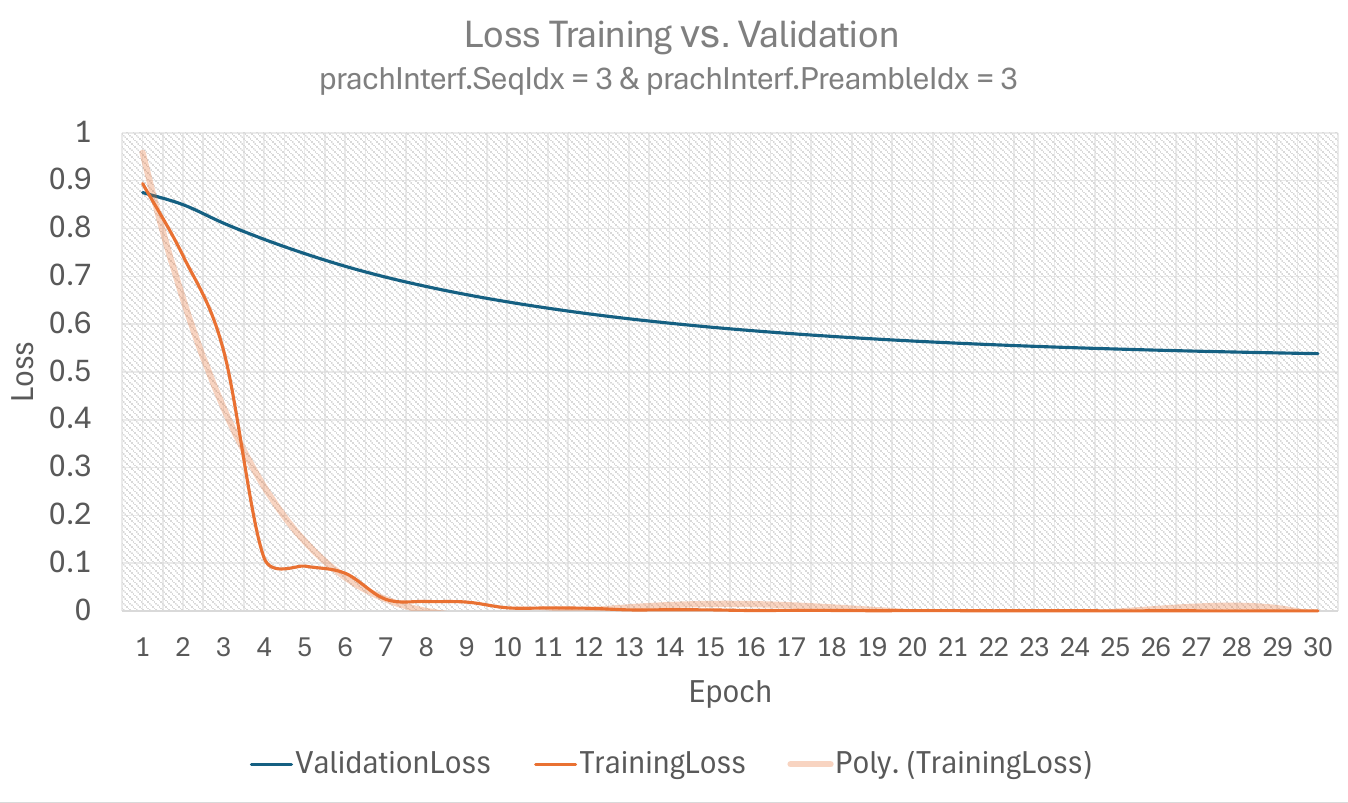}
    \caption{Training vs. Validation Loss Parameter Changes Between Training and Inference}
    \label{fig:Figure2Resultats_5}
\end{figure*}

Figures \ref{fig:Figure2Resultats_4} and \ref{fig:Figure2Resultats_5} provide an evaluation of the model's performance in terms of training and validation accuracy and loss over 30 epochs. The specificity of this analysis lies in the fact that the model was trained with \textsl{\textcolor{myColor}{prachInterf.SeqIdx = 22}} and inference was performed with \textsl{\textcolor{myColor}{prachInterf.SeqIdx = 3}}, introducing a change in conditions between training and inference.

Figure \ref{fig:Figure2Resultats_4} shows that the training accuracy increases rapidly, reaching nearly 100\% within the first few epochs and then stabilizing. In contrast, the validation accuracy increases more slowly and stabilizes around 75\%. The difference between the training and validation curves indicates overfitting, already identified in the results of Figures \ref{fig:Figure2Resultats_1} and \ref{fig:Figure2Resultats_2}. Again, the model learns the training data very well but struggles to generalize to new data, especially due to the change in \textsl{\textcolor{myColor}{SeqIdx}} between training and inference.

Figure \ref{fig:Figure2Resultats_5} shows the training loss, which decreases rapidly to almost zero values after only a few epochs. In contrast, the validation loss decreases more slowly and stabilizes around 0.3. The polynomial trend line fitted to the training loss shows a rapid decrease, confirming that the model has learned to minimize error on the training data.

The change in the value of \textsl{\textcolor{myColor}{SeqIdx}} (which translates to a change in the cell) between training and inference introduces a variation in the data characteristics, making it difficult for the model to generalize correctly. Overfitting is clearly visible with a divergence between the training and validation curves.

\begin{figure*}[htbp]
    \centering
    \includegraphics[page=6, clip, trim=0.0cm 0.0cm 0.0cm 0.0cm, width=0.90\textwidth]{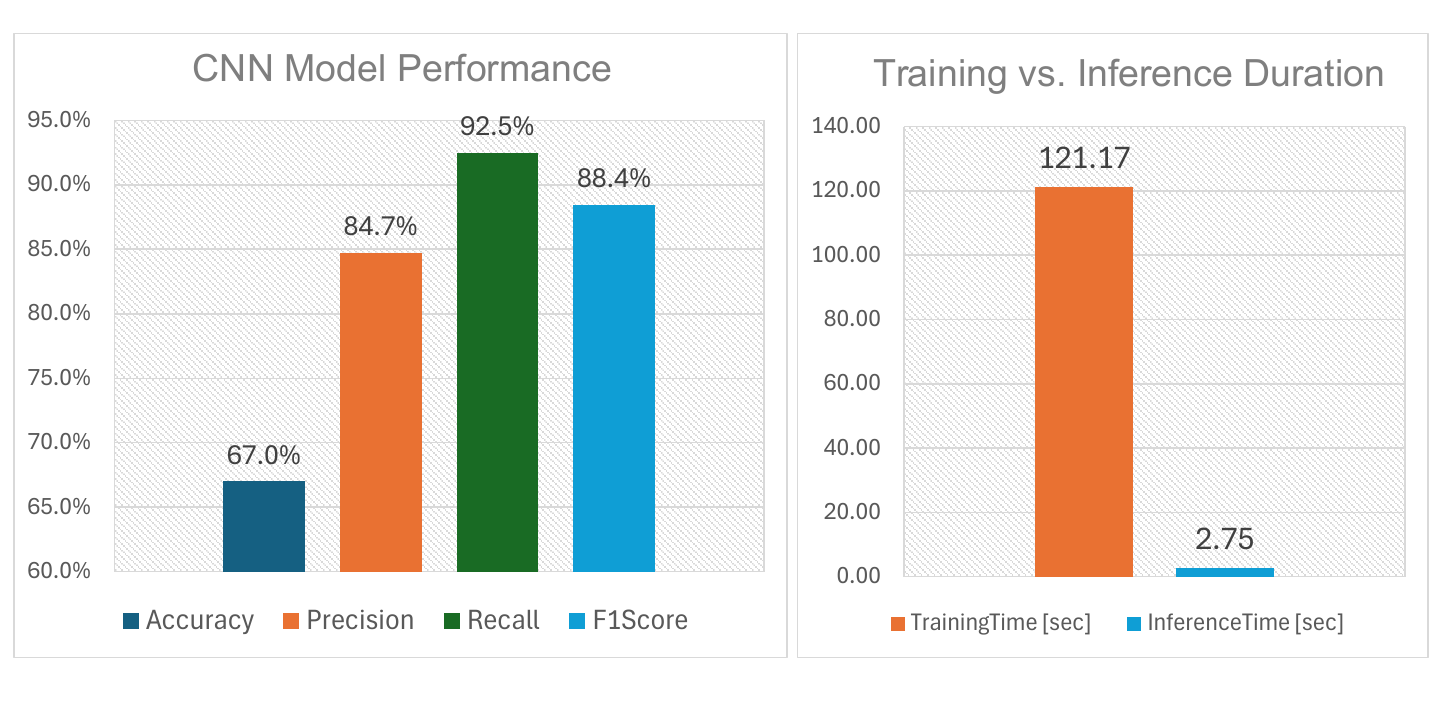}
    \caption{CNN Model Performance (Accuracy, Precision, Recall, F1-Score) and Processing Time with a Change in Parameters Between Training and Inference}
    \label{fig:Figure2Resultats_6}
\end{figure*}

The results in Figure \ref{fig:Figure2Resultats_6} indicate that the model has a very high recall, meaning it is effective at identifying interference. However, the relatively lower accuracy (67.0\%) compared to other metrics suggests that the model still makes errors, likely due to a significant number of false positives or false negatives. The high precision (84.7\%) and F1 score (88.4\%) show a good balance between precision and recall, but the gap between accuracy and other metrics highlights a difficulty in the model's generalization, as identified in the previous results. The training time is reasonable, indicating efficiency in model training, while the very fast inference time is advantageous for real-time applications.

\begin{figure*}[htbp]
    \centering
    \includegraphics[page=7, clip, trim=0.0cm 0.0cm 0.0cm 0.0cm, width=0.90\textwidth]{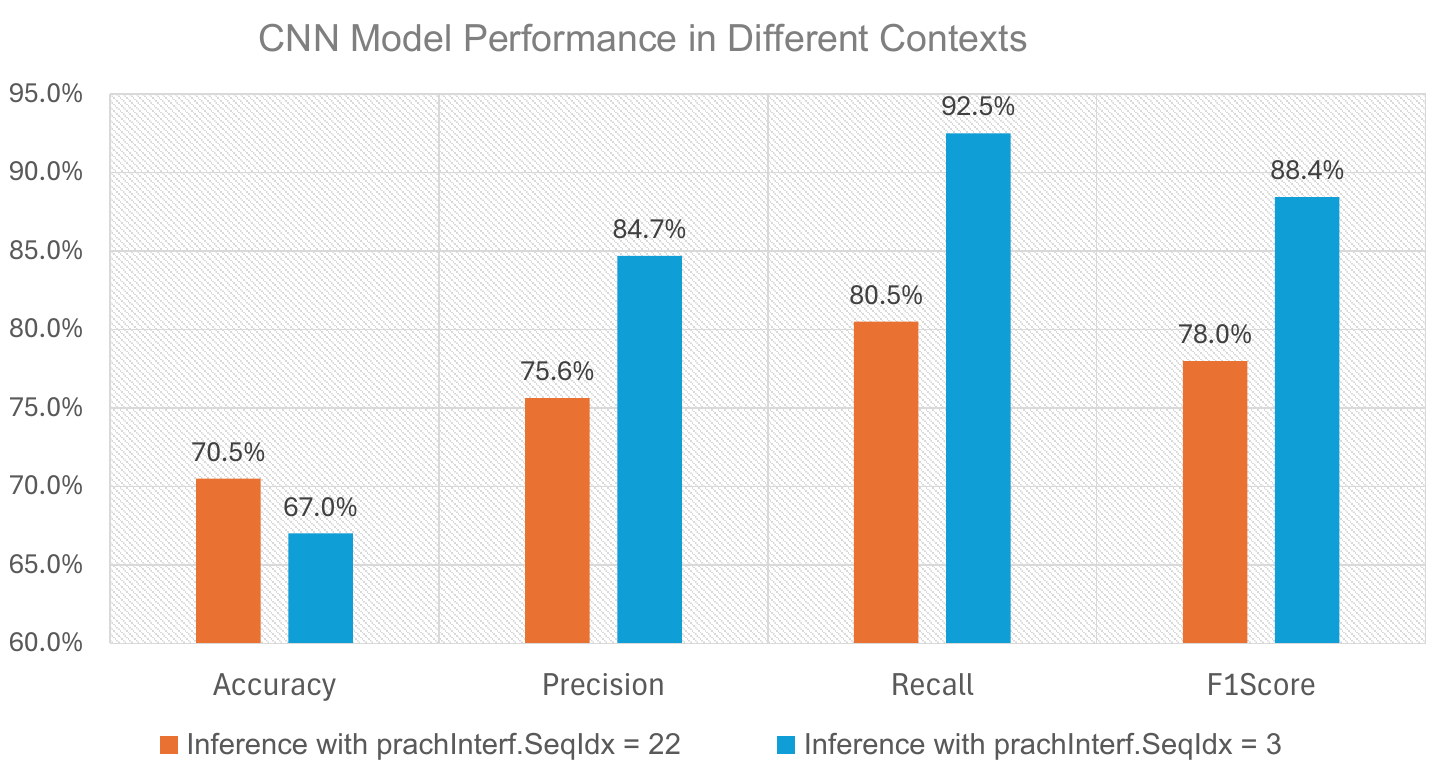}
    \caption{CNN Model Performance for PRACH Detection in the Presence of Interference with Different Sequence Indices (SeqIdx = 22 and SeqIdx = 3) in 5G-NR Networks}
    \label{fig:Figure2Resultats_7}
\end{figure*}

The results in Figure \ref{fig:Figure2Resultats_7} show a difference in the model's performance when inference is performed with SeqIdx = 22 compared to SeqIdx = 3. While accuracy is slightly better for SeqIdx = 22, the other metrics (precision, recall, F1 score) are higher for SeqIdx = 3. This might seem counterintuitive since the model is expected to perform better with the same SeqIdx seen during training. However, this observation might indicate that SeqIdx = 3 presents signal characteristics that are more easily detectable or less interfered with compared to SeqIdx = 22, which could explain the improved precision and recall metrics.

To conclude this part of the analysis, it is noted that although the model shows good overall performance, the variation in performance based on SeqIdx highlights the importance of diverse training and continuous evaluation to ensure better generalization and robustness of the CNN model.

\subsection{Limitations and Future Directions}
In this section, we examine the inherent constraints of our current study and explore opportunities for future improvement to enhance the robustness and efficiency of PRACH signal detection in highly interfered communication environments.

\subsubsection{Study Limitations}
While the results of our study are promising, showing an accuracy of 84.7\%, a recall of 92.5\%, and an F1 score of 88.4\% achieved by the CNN model for PRACH detection in the presence of interference, it is important to recognize several limitations that might influence the interpretation of our results and the generalization of our conclusions.  The  key technical, methodological and contextual limitations identified in our study are listed below:

\begin{itemize}
    \item \textsl{\textcolor{myColor}{Technical Limitations:}} Complexity of channel models, idealized simulation parameters and limited computing capacity affecting training.
    \item \textsl{\textcolor{myColor}{Methodological Limitations:}} Insufficient data sampling and evaluation methods that may not fully assess model robustness.
    \item \textsl{\textcolor{myColor}{Contextual Limitations:}} Applicability in real scenarios and the need for model adaptability to new technological standards.
\end{itemize}

These limitations highlight the importance of continuing research in this field by developing more robust models and using more diverse validation methods that can better simulate varied operational conditions. Additionally, it would be beneficial to extend future studies to include sensitivity analyses and stress tests to better understand the scope and limits of the proposed model.

\subsubsection{Future Directions}
This study opens several interesting avenues for future research and practical improvements. First, extending the training dataset to include a greater diversity of PRACH sequences (different SeqIdx) could significantly improve the model's robustness and generalization. By incorporating varied preamble sequences, the model would be better equipped to handle variations encountered in real scenarios, thereby increasing its reliability. Additionally, applying advanced regularization techniques, such as dropout, L2 regularization, and data augmentation, could help reduce the observed overfitting, thereby improving the model's overall performance on unseen validation data.


\section{Conclusion}
\label{sec:8}


In this study, we proposed and evaluated a Convolutional Neural Network (CNN)-based approach for detecting PRACH signals under interference conditions in 5G-NR networks. The proposed CNN model demonstrated a high recall rate of 92.5\%, indicating its effectiveness in identifying interference-affected signals. The precision and F1 scores were also high, at 84.7\% and 88.4\% respectively, highlighting a balanced performance in detection accuracy and reliability. The model showed a certain degree of overfitting, as evidenced by the divergence between training and validation curves. This suggests the need for further improvements, such as enhanced regularization techniques, to improve generalization to unseen data.  Performance evaluation metrics, including accuracy, precision, recall, and F1 score, indicated the model's robustness in different interference scenarios. Particularly, the variation in sequence indices (SeqIdx) between training and inference phases was found to impact performance, emphasizing the importance of diverse and continuous evaluation.

The successful application of CNNs for PRACH interference detection underlines the potential for AI/ML techniques to enhance the robustness and efficiency of 5G networks. This approach can be integrated into real-time network operations, contributing to more reliable and efficient mobile communication systems.

\printbibliography[heading=subbibliography,nottype=software,nottype=softwareversion,nottype=softwaremodule,nottype=codefragment,title={\textsc{References}}] 

@InProceedings{Dahal2022,
  author    = {Madan Dahal and Mojtaba Vaezi},
  booktitle = {56th Asilomar Conference on Signals, Systems, and Computers, {ACSSC} 2022, Pacific Grove, CA, USA, October 31 - Nov. 2, 2022},
  title     = {Deep Reinforcement Learning for Interference Management in Millimeter-Wave Networks},
  year      = {2022},
  pages     = {1064--1069},
  publisher = {{IEEE}},
  bibsource = {dblp computer science bibliography, https://dblp.org},
  comment   = {Introduction
Increasing peak data rate and user experienced data rate is among the main key performance indicators of fifth-generation (5G) and beyond cellular networks [1]. Millimeter-wave (mmWave) communications has been a promising technology to this end and has entered into 5G standardization [2]. The short wavelength of mmWave allows equipping base stations (BSs) with a large number of antennas. However, the use of mmWave comes with its unique challenges. While mmWave supports high data rates, high attenuation in mmWave incurs smaller inter-site distances resulting in higher inter-cell interference.
Inter-cell interference is a long-lasting problem in cellular networks. Interference alignment is a recent breakthrough in this field [3]–​[5]. Yet, it has not entered wireless standards due to its high complexity and huge data processing requirements. There are several sub-optimal and iterative solutions for this problem [6]. Recently deep learning has found various applications in solving difficult problems in wireless communications [7]. In particular, deep reinforcement learning (DRL) [8] has gained momentum. DRL is a subfield of deep learning that can learn to act in complex environments which are difficult to model or solve with tractable mathematical models. Recent research efforts have shown that DRL is very useful in sophisticated communications problems [9]–​[13]. Particularly, a Q-learning framework is proposed in [12] in which interference is mitigated based on the fixed power allocation. In [13] the state-action-reward-state-action algorithm was implemented for power allocation to improve spectrum access for channel selection.
In this paper, we apply DRL for interference management in multi-cell mmWave networks. Our DRL-based algorithm uses a deep Q-network (DQN) [14] to perform interference management through coordinated power allocation and beamforming. The proposed solution does not require the instantaneous channel state information which is needed for the typical signal-to-interference-plus-noise ratio (SINR) maximization problems, e.g., a brute-force method. The solution only needs the effective SINR of the user to be sent back to their BSs which helps to reduce the signaling overhead. This is meaningful progress towards viable interference management.

The performance of the proposed algorithm is compared to that of existing algorithms in terms of coverage and achievable rate [15]. Simulation results confirm that a noticeable gain in spectral efficiency can be achieved both in two- and three-cell networks. More importantly, the solution and the achieved gain scale with the number of cells.
A. Drl Overview
DRL combines reinforcement learning (RL) perception capabilities and deep learning approximation capabilities. DRL enables the agent to interact with the environment to discover what actions the agent should take to get the maximum reward. In Fig. 2, we can see how the agent interacts with an environment using a deep neural network. RL models are built using the elements explained below:

Agent: The agent acts on the environment, observes and perceives the environment, and takes action to achieve its goals.

Environment: The environment consists of the properties of the abstracted problems, which are interacted with by the agent.
B. Drl for Interference Management
We propose an algorithm that performs interference management through coordinated power allocation and beamforming. Below, we describe some of the information we used in the network.
SECTION IV.Training Setup and Simulation Results
The training setup, simulation details, performance measures and numerical results are demonstrated in this section.

A. Simulation Details and Algorithm Comparison
We have set up the experiments described below and an appropriate comparison of the algorithm to show the performance improvement of our proposed algorithm.

1) Simulation Setup
The environment parameters are given in Table I. DRL parameters are chosen according to Table II. For our simulations, we deploy three BSs, one is the serving and the others are interfering. The initial position of the UE, the initial power of the BS, and the beamforming vector are selected randomly. At least, in order to plot the effective SINR, we set the minimum SINR as γmin=−3 dB which represents the minimum SINR for any user in the cellular network. If the SINR falls below the minimum value, the episode aborts which means the call is dropped.

2) Algorithms
• Proposed Model
This is the proposed DRL algorithm explained in the previous section.

• Model 1
This model is based on training the network till it reaches the minimum value of SINR. If the network reaches the minimum value of SINR during the training, it will terminate the episode and start the process again for the next episode [20]. This is not appropriate if we want to train the model for a longer time frame not to just stop at the minimum value but to explore further.

• Model 2
This model is based on [15], where a different optimization and reward approach is used for the training of the model. It wants to maximize the sum of the SINRs rather than the product of the (1+SINRs) which is the case in our model.

• Brute Force
This algorithm is based on the exhaustive search in the space of power and beamforming vector per BS to maximize the product of the (I+SINRs). This is the best case with which our proposed algorithm is compared to.

B. Performance Measures
• Coverage
Coverage is the extent (area) to which the BS signal can be received with an acceptable quality [21], and is an important quality measure in mobile networks. We use the cumulative distribution function (CCDF) of γℓ for all cells. The SINR achieved by all users served by all BSs is recorded. The data are collected by running the experiment multiple times and then CCDF is plotted.
Conclusions
We have proposed a DQN-based interference management in multi-cell mmWave networks. In this setting, each UE experiences interference from other multi-antenna BSs. We have maximized the sum-rate achieved by all UEs. The BSs select their beamforming vector and power command from finite sets. The input features of DQN are the users' coordinates, BSs power, and beamforming vectors. The output has a sequence of interference management along with power control and beamforming that optimize the objective function. Our proposed algorithm almost reaches the spectral efficiency obtained by exhaustively researching among all possible beamforming vectors and BS powers. The complexity of the method is much lower, which makes it promising to be implemented in practice.},
  doi       = {10.1109/IEEECONF56349.2022.10052069},
  ranking   = {rank5},
  url       = {https://ieeexplore.ieee.org/document/10052069},
}

@InProceedings{Dahal2023,
  author    = {Dahal, Madan and Vaezi, Mojtaba},
  booktitle = {2023 57th Annual Conference on Information Sciences and Systems (CISS)},
  title     = {Multi-agent Deep Reinforcement Learning for Multi-Cell Interference Mitigation},
  year      = {2023},
  month     = mar,
  publisher = {IEEE},
  doi       = {10.1109/ciss56502.2023.10089622},
  url       = {https://doi.org/10.1109/ciss56502.2023.10089622},
}

@InProceedings{Xiong2018,
  author    = {Xiong, Qi and Yu, Bin and Qian, Chen and Li, Xiaojiao and Sun, Chengjun},
  booktitle = {2018 IEEE Globecom Workshops (GC Wkshps)},
  title     = {Random Access Preamble Generation and Procedure Design for 5G-NR System},
  year      = {2018},
  month     = dec,
  publisher = {IEEE},
  comment   = {1. Informations Générales**:
   L'article examine la génération de préambules d'accès aléatoire et la conception des procédures pour les systèmes 5G-NR, en se concentrant sur les différences fondamentales entre les systèmes LTE et 5G-NR, notamment en ce qui concerne l'espacement multi-sous-porteuse (SCS) et les opérations alimentées par faisceau analogique multiple.

2. **Contexte de la Recherche**:
   2.1 **Introduction**:
      L'article introduit les évolutions de la norme 5G-NR par rapport à LTE, en mettant en évidence la nécessité de concevoir de nouvelles méthodes de génération de préambules pour le canal d'accès aléatoire physique (PRACH) et les protocoles d'accès aléatoire.
   2.2 **Identification du Problème**:
      Le problème de recherche est clairement identifié comme la nécessité de concevoir des méthodes de génération de préambules et des procédures d'accès aléatoire adaptées aux spécificités du 5G-NR, en tenant compte notamment de l'espacement multi-sous-porteuse et des opérations alimentées par faisceau analogique multiple.

3. **Objectifs et Hypothèses**:
   3.1 **Objectifs de l'Étude**:
      Les objectifs de recherche sont de proposer des méthodes de génération de préambules et des procédures d'accès aléatoire adaptées aux spécificités du 5G-NR, en mettant particulièrement l'accent sur l'allocation efficace des ressources et la réduction de la latence d'accès.
   3.2 **Hypothèses**:
      Aucune hypothèse spécifique n'est mentionnée dans l'article.

4. **Méthodologie**:
   4.1 **Conception de l'Étude**:
      L'article présente une approche théorique basée sur la modélisation et la simulation pour proposer des méthodes de génération de préambules et des procédures d'accès aléatoire pour les systèmes 5G-NR.
   4.2 **Variables Mesurées**:
      Les variables mesurées incluent la performance de détection de préambules, l'impact de la transmission avec un ou plusieurs faisceaux, et l'efficacité de la réduction de la latence d'accès.

5. **Résultats**:
   5.1 **Principales Observations**:
      Les résultats montrent que les méthodes de génération de préambules et les procédures d'accès aléatoire proposées améliorent significativement la performance de détection et permettent de réduire la latence d'accès.
   5.2 **Résumé des Résultats**:
      Les méthodes proposées améliorent la détection des préambules et réduisent la latence d'accès, en particulier dans les scénarios sensibles à la latence comme les communications URLLC.

6. **Discussion**:
   6.1 **Limitations de l'Étude**:
      Les limitations de l'étude pourraient inclure des simplifications dans les modèles de simulation utilisés et des hypothèses sur les conditions de canal.
   6.2 **Comparaison avec d'Autres Travaux**:
      Les auteurs comparent leurs résultats à d'autres études dans le domaine des réseaux cellulaires et des communications par satellite.

7. **Conclusion et Implications**:
   7.1 **Conclusion Générale**:
      Les méthodes proposées pour la génération de préambules et les procédures d'accès aléatoire sont adoptées ou acceptées par les spécifications 5G-NR, ce qui souligne leur pertinence et leur efficacité.
   7.2 **Implications pour la Recherche/Pratique**:
      Les résultats de l'étude peuvent être utilisés dans la conception et le déploiement de réseaux 5G-NR pour améliorer l'efficacité des procédures d'accès aléatoire et réduire la latence.

8. **Évaluation Critique**:
   8.1 **Force de l'Étude**:
      L'article propose des solutions novatrices pour les défis spécifiques posés par les systèmes 5G-NR, et les résultats sont validés par des simulations.
   8.2 **Faiblesses et Points à Améliorer**:
      Les limitations de l'étude pourraient inclure une évaluation plus approfondie des performances dans des scénarios réels et une comparaison avec d'autres approches existantes.

9. **Références**:
   Les références citées incluent des travaux antérieurs pertinents dans le domaine des réseaux cellulaires et des communications par satellite.

10. **Liens avec Votre Recherche**:
    10.1 **Pertinence pour votre Thème**:
         Cet article est pertinent pour le thème de l'analyse de techniques AI/ML pour la gestion des interférences dans les réseaux 5G-NR, car il propose des méthodes innovantes pour améliorer l'efficacité des procédures d'accès aléatoire.
    10.2 **Contributions à votre Travail**:
         Les résultats et les idées présentés dans cet article pourraient contribuer à votre thèse en fournissant des perspectives sur la conception de méthodes AI/ML pour optimiser les performances des réseaux 5G-NR, en particulier en ce qui concerne la gestion des interférences.},
  doi       = {10.1109/glocomw.2018.8644175},
  url       = {https://ieeexplore.ieee.org/document/8644175/},
}

@Article{Zhang2019,
  author    = {Chaoyun Zhang and Paul Patras and Hamed Haddadi},
  journal   = {{IEEE} Commun. Surv. Tutorials},
  title     = {Deep Learning in Mobile and Wireless Networking: {A} Survey},
  year      = {2019},
  number    = {3},
  pages     = {2224--2287},
  volume    = {21},
  bibsource = {dblp computer science bibliography, https://dblp.org},
  doi       = {10.1109/COMST.2019.2904897},
  url       = {https://doi.org/10.1109/COMST.2019.2904897},
}

@Article{Zhang2021,
  author    = {Zhang, Qixia and Liu, Fangming and Zeng, Chaobing},
  journal   = {IEEE/ACM Transactions on Networking},
  title     = {Online Adaptive Interference-Aware VNF Deployment and Migration for 5G Network Slice},
  year      = {2021},
  issn      = {1558-2566},
  month     = oct,
  number    = {5},
  pages     = {2115--2128},
  volume    = {29},
  doi       = {10.1109/tnet.2021.3080197},
  publisher = {Institute of Electrical and Electronics Engineers (IEEE)},
  url       = {https://ieeexplore.ieee.org/document/9440734},
}

@Article{Hassan2021,
  author  = {Hassan, Ashwaq N. and Al-Chlaihawi, Sarab and Khekan, Ahlam R.},
  journal = {Indonesian Journal of Electrical Engineering and Computer Science},
  title   = {Artificial Intelligence Techniques over the Fifth Generation (5G) Mobile Networks: A Review},
  year    = {2021},
  month   = {October},
  comment = {1. Informations Générales
2. Contexte de la Recherche
Le document explore l'impact potentiel de l'intégration de l'intelligence artificielle (IA) dans les réseaux mobiles 5G, mettant en lumière la promesse de l'IA pour améliorer la gestion des réseaux, la réduction des erreurs, et la minimisation de la latence.
Les auteurs identifient les défis posés par les capacités avancées des réseaux 5G, y compris la gestion de la complexité accrue du réseau, comme des domaines où l'IA et l'apprentissage automatique (ML) peuvent offrir des solutions significatives
3. Objectifs et Hypothèses
L'objectif principal est de passer en revue et de catégoriser les méthodes utilisées par l'intelligence artificielle dans les systèmes de communication, en particulier les réseaux 5G, et de créer une vue d'ensemble des applications de l'IA dans ces réseaux
4. Méthodologie
Il s'agit d'une étude de revue de littérature, examinant diverses publications sur l'application de l'IA et du ML dans les réseaux 5G
Les variables principales incluent l'efficacité et la performance des techniques d'IA dans l'optimisation des réseaux 5G, la gestion des ressources, et l'amélioration de l'expérience utilisateur.
5. Résultats
Les techniques d'IA, y compris le ML et le deep learning (DL), sont envisagées pour jouer un rôle fondamental dans les réseaux 5G, notamment en automatisant l'infrastructure et en gérant la complexité du réseau
7. Conclusion et Implications
L'utilisation de l'IA et du ML sera cruciale pour relever les défis posés par la complexité des réseaux 5G, permettant des réseaux plus stables, une meilleure gestion des ressources, et une réduction significative de la consommation d'énergie
Le document implique que l'intégration de l'IA dans les réseaux 5G pourrait conduire à des avancées significatives dans la gestion du réseau et dans la prestation de services, offrant une nouvelle voie pour l'innovation dans la technologie des télécommunications.
8. Évaluation Critique
L'étude fournit un aperçu complet de l'application actuelle et potentielle de l'IA dans les réseaux 5G, mettant en évidence l'importance de l'IA pour l'avenir des communications mobiles
10. Liens avec Votre Recherche
Les idées et observations présentées peuvent servir de base pour explorer comment l'IA peut être appliquée pour résoudre des problèmes spécifiques dans les réseaux 5G, notamment dans la gestion des interférences et l'optimisation des ressources, ce qui pourrait être directement applicable à votre recherche sur les techniques d'IA/ML pour la gestion des interférences dans les réseaux 5G-NR.





8. Évaluation Critique

9. Références
10. Liens avec Votre Recherche
Expansion du réseau 5G
 Prévision : 100 milliards d'appareils
 Utilisateurs : 2.5 milliards de consommateurs mensuels de plus de 1 Go
Impacts et domaines d'application
Smart cities, réalité augmentée mobile, streaming vidéo 4K
Caractéristiques de la 5G
Haut débit, fiabilité accrue, faible latence
Défis pour les opérateurs
- Gestion du réseau, réduction des erreurs, minimisation de la latence
Utilisation de l'intelligence artificielle
- Solution aux défis de la 5G

3. APPLICATIONS OF AI IN FIFTH GENERATION NETWORKS
   1. Utilisation de l'IA et de l'apprentissage automatique dans les réseaux de communication 
   - Gestion efficace des ressources grâce à l'introduction des réseaux auto-organisés (SON).
   - Objectif clair des SONs : réduire les coûts opérationnels et optimiser les réseaux mobiles.
   - Vision future : planification, configuration, gestion, optimisation et dépannage facilités.
 2. Classes de performance du SON 
   - Auto-configuration
   - Auto-optimisation
   - Auto-réparation
 3. SON dans les réseaux de 5e génération 
   - Cadre général avec intégration de Big Data.
   - Trois couches de données : commune, centrale, et cellulaire.
   - Utilisation des outils de ML pour convertir les données brutes en données significatives.

 4. Réseaux hétérogènes (HetNets) 
   - Défis des HetNets en raison de l'augmentation constante des ressources.
   - Objectif du SON : réduire les coûts, optimiser la couverture, la capacité et la qualité de service (QoS).

 5. Techniques d'intelligence artificielle 
   - Diversité des méthodes inspirées par la nature, raisonnement humain, apprentissage basé sur les retours.
   - Réseaux neuronaux artificiels (ANN) comme outils fondamentaux.
   - Applications des ANN dans la modélisation, l'optimisation et la prédiction.

 6. Algorithmes évolutifs 
   - Utilisation pour résoudre les problèmes de programmation de cellules et optimiser les emplacements des nœuds dans HetNets.
   - Exemple d'utilisation des algorithmes génétiques pour maximiser la couverture et la capacité tout en réduisant les coûts.

 7. Prédiction de la QoS avec l'apprentissage automatique 
   - Application de l'apprentissage automatique pour la gestion efficace des réseaux avec l'augmentation du nombre de connexions.
   - Utilisation des réseaux neuronaux profonds pour maintenir et gérer la QoS pendant la diffusion vidéo.

 8. Sécurité et optimisation de la consommation d'énergie 
   - Utilisation de l'IA pour détecter et contrer les attaques dans les réseaux de communication.
   - Optimisation de la consommation d'énergie avec des algorithmes d'IA.

 9. Automatisation des processus et réduction des coûts 
   - Substitution ou support des processus manuels par des processus automatisés grâce à l'IA.
   - Réduction des coûts opérationnels et des erreurs humaines.},
}

@techreport{3gppLTE104,
  title = {{LTE; Evolved Universal Terrestrial Radio Access (E-UTRA); Base Station (BS) radio transmission and reception}},
  author = {{3rd Generation Partnership Project (3GPP)}},
  year = {2013},
  institution = {{3GPP}},
  number = {{TS 36.104 version 11.2.0 Release 11}},
  url = {https://www.3gpp.org/ftp/Specs/archive/36_series/36.104/},
}

@misc{matlab2022,
  author = {{The MathWorks, Inc.}},
  title = {{MATLAB} version: 9.13.0 ({R}2022b)},
  year = {2022},
  howpublished = {Accessed: January 01, 2023},
  url = {https://www.mathworks.com}
}

@MastersThesis{Yaro2022,
  author          = {Yaro, Elie Jephte},
  school          = {Université Joseph Ki-Zerbo},
  title           = {5G NR - Etude de la robustesse et de la résistance aux interférences en voie montante},
  year            = {2022},
  address         = {Ouagadougou, Burkina Faso},
  note            = {Mémoire de Master, Unité de Formation et de Recherche en Sciences Exactes et Appliquées (UFR-SEA), Département d'Informatique},
  academic_year   = {2021--2022},
  domain          = {Sciences et Technologies},
  encadrants      = {Dr Désiré GUEL & Dr P. Justin KOURAOGO},
  encadrement_par = {Dr Désiré GUEL & Dr P. Justin KOURAOGO},
  examinateur     = {Dr Didier BASSOLE},
  mention         = {Informatique},
  option          = {Systèmes, Réseaux et Télécommunication (SRT)},
  president       = {Pr Oumarou SIE},
  ranking         = {rank5},
  rapporteur      = {Dr K. Kisito KABORE},
  superviseur     = {Pr Oumarou SIE},
}

@MastersThesis{Boumaza2020,
  author     = {Boumaza, Amel and Mogdad, Yamina},
  school     = {Université Kasdi Merbah Ouargla},
  title      = {Mémoire de fin d'étude},
  year       = {2020},
  month      = {October},
  note       = {Soutenu en octobre 2020 devant le jury composé de: Mr. CHEBARA Fouad (Président, UKM Ouargla), Mr. MOAD Mouhamed Esayeh (Examinateur, UKM Ouargla), Mr. AOUNALLAH Naceur (Encadreur, UKM Ouargla)},
  comment    = {1. Informations Générales
2. Contexte de la Recherche
2.1 Introduction 
L'introduction générale met en lumière l'évolution actuelle vers la 5G, soulignant les avantages significatifs que cette technologie promet d'apporter, tels qu'une vitesse de téléchargement substantiellement accrue, une meilleure connectivité, et l'intégration avec l'Internet des objets (IoT) pour réaliser un monde hyper-connecté​
2.2 Introduction 
Le problème de recherche abordé consiste à trouver des solutions efficaces pour gérer la complexité croissante des réseaux 5G, où l'apprentissage profond est envisagé comme une solution potentielle pour surmonter ces défis, en particulier dans le contexte du beamforming hybride et de l'estimation de canal.

3. Objectifs et Hypothèses
Objectifs de l'Étude 
Quels sont les objectifs de recherche définis par les auteurs ?
Les objectifs de l'étude se concentrent sur l'évaluation de l'efficacité des techniques d'apprentissage profond dans l'amélioration des performances des systèmes 5G, en particulier pour ce qui est de la formation de faisceaux, de l'allocation de puissance, et de la suppression du brouillage.
4. Méthodologie
Conception de l'Étude
Type d'étude (expérimentale, observationnelle, modélisation, etc.)
L'étude applique des méthodes d'apprentissage profond, notamment le deep learning, pour aborder des problèmes spécifiques dans les systèmes 5G à onde millimétrique, ce qui implique une approche expérimentale fondée sur la simulation et l'analyse de performance.
Variables Mesurées
Les variables incluent la performance du beamforming, l'efficacité de l'allocation de puissance, et la capacité à supprimer le signal de brouillage en utilisant des techniques d'apprentissage profond.
5. Résultats
Principales Observations :
Résumé des résultats les plus significatifs.
L'application de l'apprentissage profond améliore le compromis complexité-performances pour l'allocation de puissance.
Les techniques d'apprentissage profond permettent une suppression efficace du signal de brouillage.
Le beamforming hybride assisté par apprentissage profond offre un équilibre entre les coûts et les performances, en utilisant des composants analogiques et numériques.
6. Discussion
Comparaison avec d'Autres Travaux :
Les auteurs comparent-ils leurs résultats à d'autres études ?
L'étude compare le beamforming assisté par apprentissage profond à des méthodes conventionnelles, démontrant un avantage de performances significatif.
7. Conclusion et Implications
Conclusion Générale :
Résumé des principales conclusions de l'article.
L'apprentissage profond représente une solution prometteuse pour relever les défis des systèmes 5G à onde millimétrique, en améliorant la performance et l'efficacité des communications sans fil.
Implications pour la Recherche/Pratique :
Comment les résultats peuvent-ils être appliqués dans d'autres contextes ?
L'étude met en évidence l'importance croissante de l'apprentissage profond dans le développement des technologies de communication de prochaine génération, offrant des pistes pour de futures recherches et applications pratiques.
8. Évaluation Critique
Force de l'Étude :
Points forts de l’article.
L'étude démontre l'efficacité et le potentiel des techniques d'apprentissage profond dans l'amélioration significative des performances des réseaux 5G.
10. Liens avec Votre Recherche
Pertinence pour votre  thème :
Comment cet article est-il lié à votre recherche actuelle ou future ?
Cette étude est directement pertinente pour l'analyse des techniques d'IA/ML pour la gestion des interférences dans les réseaux 5G-NR, offrant un aperçu des applications pratiques de l'apprentissage profond.},
  domain     = {Sciences et Technologie},
  filiere    = {Télécommunication},
  ranking    = {rank5},
  readstatus = {read},
  specialite = {Systèmes de Télécommunication},
  url        = {URL_de_votre_memoire},
}

@InProceedings{Fang2022,
  author    = {Fang, Ruijie and Chen, Huamin and Liu, Wei},
  booktitle = {2022 2nd International Conference on Frontiers of Electronics, Information and Computation Technologies (ICFEICT)},
  title     = {Deep Learning-Based PRACH Detection Algorithm Design and Simulation},
  year      = {2022},
  month     = aug,
  publisher = {IEEE},
  comment   = {1. Informations Générales:
   L'article propose un algorithme de détection de PRACH basé sur l'apprentissage en profondeur pour les réseaux 5G-NR. Il utilise un réseau neuronal entièrement connecté pour réduire la complexité du traitement du signal sans perte de taux de détection.

2. Contexte de la Recherche:
   L'introduction fournit un contexte clair sur l'importance de la technologie 5G-NR et des canaux PRACH dans les réseaux sans fil. Le problème de la détection de PRACH est identifié comme un défi clé dans le traitement des signaux sans fil.

3. Objectifs et Hypothèses:
   Les objectifs de l'étude sont clairement définis : proposer un algorithme de détection de PRACH basé sur l'apprentissage en profondeur pour améliorer la précision et réduire la complexité du traitement du signal. Aucune hypothèse spécifique n'est mentionnée.

4. Méthodologie:
   L'étude utilise un réseau neuronal entièrement connecté pour la détection de PRACH et la détection du timing advance (TA). Les variables mesurées incluent les performances de détection et de précision de l'algorithme.

5. Résultats:
Les résultats de l'étude montrent une précision élevée dans la détection des séquences PRACH, avec des taux de précision de 96% et 99% pour les longueurs de séquence testées (139 et 839 respectivement). Ces résultats sont obtenus en utilisant un algorithme basé sur un réseau neuronal entièrement connecté. Les performances sont évaluées à l'aide de données d'entraînement et de test, et les graphiques présentés dans l'article montrent l'évolution de la précision et de la perte au fil des époques d'entraînement.

Pour la simulation avec une longueur de séquence PRACH de 139, environ 95% des données sont utilisées pour l'entraînement et 5% pour les tests. Avec 200 époques d'entraînement, la précision de détection atteint 96%. De même, pour la simulation avec une longueur de séquence PRACH de 839, la précision de détection atteint 99% avec 150 époques d'entraînement.

Ces résultats indiquent que l'algorithme proposé est efficace pour la détection de PRACH, ce qui suggère qu'il pourrait être utilisé dans des applications pratiques pour améliorer les performances des réseaux 5G-NR en termes de précision de détection et de réduction de la complexité du traitement du signal.
6. Discussion:
   Les limitations de l'étude incluent la nécessité de convertir les signaux du domaine temporel au domaine fréquentiel avant la simulation, ainsi que la complexité des modèles pour les signaux du domaine temporel.

7. Conclusion et Implications:
   L'article conclut que l'algorithme proposé est efficace pour la détection de PRACH avec une précision élevée. Les implications pratiques incluent son application potentielle dans les réseaux sans fil 5G-NR.

8. Évaluation Critique:
   Points forts: Méthodologie clairement décrite, résultats significatifs avec des performances de détection élevées.
   Faiblesses et Points à Améliorer: Nécessité de convertir les signaux du domaine temporel au domaine fréquentiel, les limitations de la complexité des modèles pour les signaux du domaine temporel devraient être abordées.

9. Références:
   L'article fait référence à des recherches antérieures sur la détection de PRACH et l'apprentissage en profondeur dans les réseaux sans fil.

10. Liens avec Votre Recherche:
    Pertinence pour votre thème: L'article examine les techniques d'IA/ML pour la gestion des interférences dans les réseaux 5G-NR, ce qui correspond à votre thème.
    Contributions à votre travail: Les résultats et les idées de cet article pourraient être utiles pour votre recherche sur l'optimisation des performances des réseaux 5G-NR en utilisant l'apprentissage en profondeur pour la détection et la gestion des interférences.

Cette analyse fournit une évaluation critique de l'article en se concentrant sur ses aspects méthodologiques, ses résultats et ses implications pour la recherche future dans le domaine des réseaux 5G-NR.},
  doi       = {10.1109/icfeict57213.2022.00094},
  ranking   = {rank5},
  url       = {https://ieeexplore.ieee.org/document/9951528},
}

@Article{Zhang2018,
  author        = {Chaoyun Zhang and Paul Patras and Hamed Haddadi},
  journal       = {CoRR},
  title         = {Deep Learning in Mobile and Wireless Networking: {A} Survey},
  year          = {2018},
  volume        = {abs/1803.04311},
  archiveprefix = {arXiv},
  bibsource     = {dblp computer science bibliography, https://dblp.org},
  comment       = {2. Contexte de la Recherche
2.1 Introduction 
Brève description du contexte dans lequel l'article s'inscrit.
L’article étudie les difficultés engendrées par la croissance rapide des flux de trafic mobile et des demandes de services d’application mobile, en soulignant le potentiel du deep learning pour améliorer la gestion des réseaux mobiles et sans fil.
2.2 Introduction 
Identification du problème de recherche ou de la question posée.
De quelle manière peut-on exploiter le deep Learning afin d’améliorer les performances et la gestion des réseaux mobiles et sans fil en réponse aux demandes croissantes et à la complexité croissante des environnements de réseau ?
3. Objectifs et Hypothèses
Objectifs de l'Étude 
Quels sont les objectifs de recherche définis par les auteurs ?
Élaborer une analyse exhaustive des contributions récentes et des avancées dans l’application du deep learning aux réseaux mobiles et sans fil, et repérer les défis et les perspectives à venir de la recherche dans ce domaine.
Hypothèses (si applicables) 
Les auteurs énoncent-ils des hypothèses à tester ?
La capacité du deep learning à apprendre des modèles complexes et à faire des inférences à partir de grandes quantités de données permet d’améliorer considérablement la gestion et l’efficacité des réseaux mobiles et sans fil.
4. Méthodologie
Conception de l'Étude
Revue de littérature - Cet article réunit et examine les études existantes concernant l’utilisation du Deep Learning sur les réseaux mobiles et sans fil.
Type d'étude (expérimentale, observationnelle, modélisation, etc.)
Variables Mesurées
Quelles variables ont été mesurées ou manipulées ?
5. Résultats
Principales Observations :
Les avantages du deep learning pour la gestion des réseaux sont importants, notamment en ce qui concerne l’optimisation des ressources réseau, la sécurité, la gestion du trafic et la prédiction de la demande.
Résumé des résultats les plus significatifs.
6. Discussion
Limitations de l'Étude :
Quelles sont les limitations mentionnées par les auteurs ?
Comparaison avec d'Autres Travaux :
Les auteurs comparent-ils leurs résultats à d'autres études ?
7. Conclusion et Implications
Conclusion Générale :
L’utilisation du deep learning présente des perspectives prometteuses pour résoudre de nombreux problèmes complexes dans les réseaux mobiles et sans fil, mais elle requiert des études supplémentaires afin de relever les Défis liés à l’évolutivité, à la sécurité et à l’interopérabilité.
Résumé des principales conclusions de l'article.
Implications pour la Recherche/Pratique :
Comment les résultats peuvent-ils être appliqués dans d'autres contextes ?
8. Évaluation Critique
Force de l'Étude :
Points forts de l’article.
Faiblesses et Points à Améliorer :
Limitations ou domaines d'amélioration identifiés.
Expose de manière exhaustive les applications actuelles du Deep Learning dans les réseaux mobiles et rencontrèrent en évidence les avancées potentielles à venir.
Il n’existe pas suffisamment d’études de cas concrètes ou d’exemples concrets d’implémentations qui illustrent les Bénéfices du deep learning dans des scénarios de réseau particuliers.
9. Références
Autres Travaux Cités :
Références importantes citées dans l'article.
10. Liens avec Votre Recherche
Pertinence pour votre  thème :
Comment cet article est-il lié à votre recherche actuelle ou future ?
Contributions à votre travail : 
En quoi les résultats ou les idées de cet article peuvent-ils contribuer à votre thèse ?},
  doi           = {10.48550/arxiv.1803.04311},
  eprint        = {1803.04311},
  ranking       = {rank5},
  readstatus    = {read},
  url           = {http://arxiv.org/abs/1803.04311},
}

@Article{Itu2020,
  author   = {Zhao, Houlin},
  journal  = {ITU News Magazine},
  title    = {AI and Machine Learning in 5G—The ITU Challenge 2020},
  year     = {2020},
  volume   = {No. 05},
}

@TechReport{Misc2021,
  title     = {5G‐NR Radio Interface – Radio Access Procedure},
  year      = {2021},
  month     = aug,
  doi       = {10.1002/9781119851288.ch9},
  isbn      = {9781119851288},
  journal   = {NG‐RAN and 5G‐NR},
  pages     = {237-261},
  publisher = {Wiley},
  url       = {https://onlinelibrary.wiley.com/doi/10.1002/9781119851288.ch9},
}

@Misc{Launay2021,
  author    = {Frédéric Launay},
  title     = {NG‐RAN and 5G‐NR},
  year      = {2021},
  doi       = {10.1002/9781119851288},
  publisher = {Wiley},
  subtitle  = {5G Radio Access Network and Radio Interface},
  url       = {https://doi.org/10.1002/9781119851288},
}

@Misc{Cox2021,
  author    = {Christopher Cox},
  title     = {An Introduction to 5G},
  year      = {2021},
  doi       = {10.1002/9781119602682},
  publisher = {Wiley},
  subtitle  = {The New Radio, 5G Network and Beyond},
  url       = {https://onlinelibrary.wiley.com/doi/book/10.1002/9781119602682},
}

@Misc{Moltchanov2022,
  author = {D. Moltchanov and E. Sopin and Vyacheslav Begishev and Andrey K. Samuylov and Y. Koucheryavy and K. Samouylov},
  title  = {A Tutorial on Mathematical Modeling of 5G/6G Millimeter Wave and Terahertz Cellular Systems},
  year   = {2022},
}

@TechReport{intelAI5G,
  author      = {Hong, Huisuk (Kevin) and Ruan, Leifeng and Zhang, Tong},
  institution = {Intel},
  title       = {AI in the 5G Network: Six Questions You Weren’t Supposed to Ask},
  year        = {Year},
  note        = {Communications Service Providers AI-based Network Optimization},
  type        = {White Paper},
}

@techreport{3gpp2018,
  author = "{3GPP}",
  title = "{Physical Channels and Modulation}",
  institution = "3rd Generation Partnership Project (3GPP)",
  type = "Technical Specification",
  number = "TS 38.211",
  year = 2018,
  version = "15.3.0",
  url = "https://www.3gpp.org/ftp/Specs/archive/38_series/38.211/38211-f30.zip"
}

@techreport{ts38113,
  author = "{3GPP}",
  title = "{NR; Physical Layer Procedures for Control}",
  institution = "3rd Generation Partnership Project (3GPP)",
  type = "Technical Specification",
  number = "TS 38.213",
  year = 2018,
  version = "15.2.0",
  url = "https://www.3gpp.org/ftp//Specs/archive/38_series/38.213/38213-f20.zip"
}

@article{zadoff1963,
  author = "R. Frank",
  title = "{Polyphase codes with good nonperiodic correlation properties}",
  journal = "IEEE Transactions on Information Theory",
  volume = 9,
  number = 1,
  pages = "43--45",
  year = 1963,
  doi = "10.1109/TIT.1963.1057827",
  url = "https://ieeexplore.ieee.org/document/1057827"
}

@misc{2401.12803v1,
  author = "R. Singh and A. K. Yerrapragada and J. K. S and R. K. Ganti",
  title = "{Enhancements for 5G NR PRACH Reception: An AI/ML Approach}",
  year = 2024,
  doi = "10.48550/arXiv.2401.12803",
  url = "https://arxiv.org/abs/2401.12803v1"
}

@inproceedings{pham2019,
  author = "T. A. Pham and B. T. Le",
  title = "{A proposed preamble detection algorithm for 5G-PRACH}",
  booktitle = "2019 International Conference on Advanced Technologies for Communications (ATC)",
  organization = "IEEE",
  year = 2019,
  pages = "210--214",
  doi = "10.1109/ATC.2019.8924546",
  url = "https://ieeexplore.ieee.org/document/8924546"
}

@inproceedings{kamata2021,
  author = "K. Kamata and M. Sawahashi and Y. Kishiyama",
  title = "{Detection Probability of PRACH Preamble for NR in 3GPP TDL Channel Models}",
  booktitle = "2021 IEEE VTS 17th Asia Pacific Wireless Communications Symposium (APWCS)",
  organization = "IEEE",
  year = 2021,
  pages = "1--5",
  doi = "10.1109/APWCS53836.2021.9646048",
  url = "https://ieeexplore.ieee.org/document/9646048"
}

@article{jang2021,
  author = "H. S. Jang and H. Lee and T. Q. S. Quek and H. Shin",
  title = "{Deep Learning-Based Cellular Random Access Framework}",
  journal = "IEEE Transactions on Wireless Communications",
  volume = 20,
  number = 11,
  pages = "7503--7518",
  year = 2021,
  doi = "10.1109/TWC.2021.3093808",
  url = "https://ieeexplore.ieee.org/document/9459492"
}

@article{zehra2022,
  author = "S. S. Zehra and M. Magarini and R. Qureshi and S. M. N. Mustafa and F. Farooq",
  title = "{Proactive approach for preamble detection in 5g-nr prach using supervised machine learning and ensemble model}",
  journal = "Scientific Reports",
  volume = 12,
  number = 1,
  pages = 8378,
  year = 2022,
  doi = "10.1038/s41598-022-12429-8",
  url = "https://www.nature.com/articles/s41598-022-12429-8"
}

@inproceedings{Guel2023,
  author    = {Désiré Guel and Pegdwindé Justin Kouraogo and Boureima Zerbo and Elie Jephte Yaro},
  title     = {5G-NR PRACH Detection Performance Optimization in Context of Intra/Inter-Cell Interference},
  booktitle = {Proceedings of the 6th Computer Science Research Days, JRI 2023},
  year      = {2023},
  address   = {Ouagadougou, Burkina Faso},
  month     = {December 18-20},
  url       = {https://eudl.eu/doi/10.4108/eai.18-12-2023.2348128},
  doi       = {10.4108/eai.18-12-2023.2348128},
  comment    = {**Analyse critique de l'article "5G-NR PRACH Detection Performance Optimization in Context of Intra/Inter-Cell Interference"**

1. **Informations Générales**
   L'article explore l'optimisation de la performance de détection PRACH dans les réseaux 5G-NR en tenant compte des interférences intra-cellulaires et inter-cellulaires. Il utilise des simulations MATLAB pour évaluer les performances et propose des stratégies d'amélioration.

2. **Contexte de la Recherche**
   2.1 **Introduction** : L'article situe le contexte de la recherche en soulignant l'importance de la gestion des interférences dans les réseaux 5G-NR pour assurer des performances robustes.
   2.2 **Introduction** : Le problème de recherche est clairement identifié comme l'impact des paramètres de configuration UE/cellule sur les performances de détection PRACH dans des conditions d'interférences.

3. **Objectifs et Hypothèses**
   3.1 Les objectifs de l'étude sont de comprendre l'impact des paramètres de configuration UE/cellule sur les performances de détection PRACH et de proposer des stratégies d'optimisation.
   3.2 Aucune hypothèse spécifique n'est clairement énoncée dans l'article.

4. **Méthodologie**
   4.1 L'étude utilise des simulations MATLAB pour évaluer les performances de détection PRACH dans différents scénarios d'interférence.
   4.2 Les variables mesurées incluent les performances de détection PRACH, les niveaux d'interférence intra-cellulaire et inter-cellulaire, ainsi que les paramètres de configuration UE/cellule.

5. **Résultats**
   5.1 Les principales observations incluent l'impact négatif des niveaux d'interférence intra-cellulaire et inter-cellulaire sur les performances de détection PRACH.
Observation 1: Les performances de détection PRACH, telles que la probabilité de détection du préambule, diminuent à mesure que le niveau d'interférence intra-cellulaire augmente. Cela signifie que lorsque plusieurs UE tentent d'accéder simultanément au réseau dans la même cellule, le niveau d'interférence augmente, ce qui peut compromettre la capacité du système à détecter correctement les préambules PRACH et à établir des connexions.
Observation 2: De manière similaire, les performances de détection PRACH diminuent également avec l'augmentation du niveau d'interférence inter-cellulaire. Cela se produit lorsque les UE dans des cellules voisines interfèrent les unes avec les autres, entraînant une diminution de la qualité du signal et une capacité réduite à détecter correctement les préambules PRACH.
   5.2 Les résultats mettent en évidence la sensibilité des performances de détection PRACH aux niveaux d'interférence et aux paramètres de configuration UE/cellule.

6. **Discussion**
   6.1 Les limitations de l'étude pourraient inclure des simplifications dans les modèles de simulation et des scénarios d'interférence qui pourraient ne pas refléter fidèlement les conditions réelles.
   6.2 Aucune comparaison directe avec d'autres études n'est présentée dans l'article.

7. **Conclusion et Implications**
   7.1 La conclusion résume les principales observations sur l'impact des interférences et de la configuration UE/cellule sur les performances de détection PRACH.
   7.2 Les implications pratiques incluent la nécessité de développer des stratégies d'optimisation pour gérer les interférences et améliorer les performances des réseaux 5G-NR.

8. **Évaluation Critique**
   8.1 Les points forts incluent une analyse approfondie des performances de détection PRACH dans des scénarios d'interférence variés et des propositions de stratégies d'optimisation.
   8.2 Les faiblesses pourraient inclure le manque de validation des résultats avec des données réelles et des hypothèses non explicitées.

9. **Références**
   Les références citées dans l'article fournissent un contexte académique solide pour la recherche menée.

10. **Liens avec Votre Recherche**
    10.1 Cet article est pertinent pour le thème de l'analyse de techniques AI/ML pour la gestion des interférences dans les réseaux 5G-NR, car il fournit des informations sur l'impact des interférences sur les performances de détection PRACH.
    10.2 Les résultats et les idées présentés dans cet article peuvent être utilisés pour orienter les recherches futures sur l'optimisation des performances des réseaux 5G-NR en présence d'interférences.},
  journal    = {JRI},
  ranking    = {rank5},
  readstatus = {read},
}

@book{AMukherjee2019,
  author    = {Amitav Mukherjee},
  title     = {5G New Radio: Beyond Mobile Broadband},
  year      = {2019},
  isbn      = {9781630816421, 1630816426},
  publisher = {Artech House},
  address   = {Norwood, MA},
  month     = {October 31}
}

@manual{3GPPTS36211V890,
  title     = {3GPP TS 36.211 V8.9.0 Evolved Universal Terrestrial Radio Access (EUTRA); Physical Channels and Modulation},
  year      = {2009},
  pages     = {7--44},
  note      = {3GPP TS 36.211 V8.9.0},
  url       = {https://www.3gpp.org/ftp/Specs/archive/36_series/36.211/36211-890.pdf},
  doi       = {10.1109/IEEESTD.2009.5305759}
}

@article{kattenborn2021review,
  title={Review on Convolutional Neural Networks (CNN) in vegetation remote sensing},
  author={Kattenborn, Teja and Leitloff, Jens and Schiefer, Felix and Hinz, Stefan},
  journal={ISPRS journal of photogrammetry and remote sensing},
  volume={173},
  pages={24--49},
  year={2021},
  publisher={Elsevier}
}

@inproceedings{cheraghinia2024explainable,
  title={Explainable AI (XAI) for Wireless Communications: UWB Radar for Zone-based Obstacle Detection},
  author={Cheraghinia, M. and Poorter, E. D. and Shahid, A.},
  booktitle={2024 IEEE International Conference on Localization and GNSS (ICL-GNSS)},
  year={2024},
  pages={1-6},
  doi={10.1109/ICL-GNSS60721.2024.10578362},
  url={https://dx.doi.org/10.1109/ICL-GNSS60721.2024.10578362}
}

@article{raj2024dynamic,
  title={Dynamic Trusted Cross-layer IDS for Secured Communications in Wireless Networks using Routing Algorithm and FT-CNN},
  author={Raj, K. R. M. and Katiravan, J.},
  journal={Journal of Intelligent \& Fuzzy Systems},
  year={2024},
  doi={10.3233/jifs-233275},
  url={https://dx.doi.org/10.3233/jifs-233275}
}

@article{wang2023study,
  title={Study of Interference Detection of Rail Transit Wireless Communication System Based on Fourth-Order Cyclic Cumulant},
  author={Wang, P. and Yao, J. and Pu, Y. and Zhang, S. and Wen, L.},
  journal={Sensors},
  volume={23},
  number={19},
  year={2023},
  doi={10.3390/s23198291},
  url={https://dx.doi.org/10.3390/s23198291}
}

@article{madasamy2023novel,
  title={A Novel Back-Propagation Neural Network for Intelligent Cyber-Physical Systems for Wireless Communications},
  author={Madasamy, N. S. and Eldho, K. J. and Senthilnathan, T. and Deny, J.},
  journal={Transactions of the Institute of Measurement and Control},
  year={2023},
  doi={10.1080/03772063.2023.2173669},
  url={https://dx.doi.org/10.1080/03772063.2023.2173669}
}

@inproceedings{akila2023forecasting,
  title={Forecasting Naturally Occurring Forest Fires using AI and Machine Learning in Wireless Sensor Networks},
  author={Akila, T. and M, M. and Vasanth, P. and Jancee, B. V.},
  booktitle={2023 IEEE International Conference on Computing, Communication, and Networking Technologies (ICCCNT)},
  year={2023},
  pages={1-6},
  doi={10.1109/ICIRCA57980.2023.10220860},
  url={https://dx.doi.org/10.1109/ICIRCA57980.2023.10220860}
}

@article{massimi2023artificial,
  title={Artificial Intelligence-based Hidden Communications Detection in Wireless Networks},
  author={Massimi, F. and Benedetto, F.},
  journal={IEEE Transactions on Wireless Communications},
  year={2023},
  doi={10.1109/TSP59544.2023.10197831},
  url={https://dx.doi.org/10.1109/TSP59544.2023.10197831}
}

@inproceedings{chen2023reservoir,
  title={Reservoir Computing for Symbol Detection of Optical Wireless Scattering Communications},
  author={Chen, H. and Natsuaki, R. and Hirose, A.},
  booktitle={2023 International Joint Conference on Neural Networks (IJCNN)},
  year={2023},
  pages={1-7},
  doi={10.1109/IJCNN54540.2023.10191220},
  url={https://dx.doi.org/10.1109/IJCNN54540.2023.10191220}
}

@inproceedings{sagar2023wireless,
  title={Wireless Sensor Network-based Intrusion Detection Technique using Deep Learning Approach of CNN-GRU},
  author={Sagar, A. and Anushkannan, N. K. and Indumathi, G. and Muralidhar, N. V. and Dhamotharan, K. A. and Malini, P.},
  booktitle={2023 IEEE International Conference on Computing and Network Communications (CoCoNet)},
  year={2023},
  pages={1-6},
  doi={10.1109/ICCES57224.2023.10192844},
  url={https://dx.doi.org/10.1109/ICCES57224.2023.10192844}
}

@article{noori2023detection,
  title={Detection and Reporting of Wireless Channel Congestion and Interference in Connected Vehicle Networks},
  author={Noori, H. and Shen, R. and Khandakar, A. and de Geuser, L. and Michelson, D.},
  journal={Transportation Research Record},
  volume={2677},
  number={7},
  pages={1-12},
  year={2023},
  doi={10.1177/03611981221112421},
  url={https://dx.doi.org/10.1177/03611981221112421}
}

@article{li2023intelligent,
  title={Intelligent metasurface system for automatic tracking of moving targets and wireless communications based on computer vision},
  author={Li, W. and Ma, Q. and Liu, C. and Zhang, Y. and Wu, X. and Wang, J. and Gao, S. and Qiu, T. and Liu, T. and Xiao, Q. and Wei, J. and Gu, T. and Zhou, Z. and Li, F. and Cheng, Q. and Li, L. and Tang, W. and Cui, T.},
  journal={Nature Communications},
  volume={14},
  pages={1-10},
  year={2023},
  doi={10.1038/s41467-023-36645-3},
  url={https://dx.doi.org/10.1038/s41467-023-36645-3}
}

@inproceedings{robinson2023narrowband,
  title={Narrowband Interference Detection via Deep Learning},
  author={Robinson, C. P. and Uvaydov, D. and D’oro, S. and Melodia, T.},
  booktitle={2023 IEEE International Conference on Communications (ICC)},
  year={2023},
  pages={1-6},
  doi={10.1109/ICC45041.2023.10278618},
  url={https://dx.doi.org/10.1109/ICC45041.2023.10278618}
}

@article{zhang2022deep,


  title={Deep Learning-Based Signal Detection for Underwater Acoustic OTFS Communication},
  author={Zhang, Y. and Zhang, S. and Wang, B. and Liu, Y. and Bai, W. and Shen, X.},
  journal={Journal of Marine Science and Engineering},
  volume={10},
  number={12},
  year={2022},
  doi={10.3390/jmse10121920},
  url={https://dx.doi.org/10.3390/jmse10121920}
}

@inproceedings{hussain2022jamming,
  title={Jamming Detection in IoT Wireless Networks: An Edge-AI Based Approach},
  author={Hussain, A. M. and Abughanam, N. and Qadir, J. and Mohamed, A.},
  booktitle={2022 IEEE International Conference on Artificial Intelligence for Industries (AI4I)},
  year={2022},
  pages={1-6},
  doi={10.1109/AI4I55679.2022.9855170},
  url={https://dx.doi.org/10.1109/AI4I55679.2022.9855170}
}

@article{he2022novel,
  title={A Novel Approach Based on Generative Adversarial Network for Interference Detection in Wireless Communications},
  author={He, S. and Zhu, L. and Yao, C. and Zeng, W. and Qin, Z.},
  journal={Journal of Electrical and Computer Engineering},
  volume={2022},
  pages={1-9},
  year={2022},
  doi={10.1155/2022/7050573},
  url={https://dx.doi.org/10.1155/2022/7050573}
}
\thispagestyle{fancy}

\end{document}